\documentclass[aps,prd,twocolumn,amsmath,amssymb,showpacs,footnote,reprint]{revtex4-1}

\usepackage{graphicx}
\usepackage{dcolumn}
\usepackage{bm}
\usepackage{appendix}
\usepackage{enumitem}

\usepackage{epstopdf} 

\usepackage[colorlinks,hyperindex]{hyperref}  
\hypersetup
    {
        colorlinks,%
        citecolor=blue,%
        linkcolor=blue,%
        urlcolor=blue,%
    }

\usepackage{subfigure}

\begin{document}

\title{Waveforms produced by a scalar point particle plunging\\ into a Schwarzschild black hole: \\
Excitation of quasinormal modes and quasibound states.}

\author{Yves D\'ecanini}
\email{decanini@univ-corse.fr}

\affiliation{Equipe Physique
Th\'eorique - Projet COMPA, \\ SPE, UMR 6134 du CNRS
et de l'Universit\'e de Corse,\\
Universit\'e de Corse, BP 52, F-20250 Corte,
France}

\author{Antoine Folacci}
\email{folacci@univ-corse.fr}

\affiliation{Equipe Physique
Th\'eorique - Projet COMPA, \\ SPE, UMR 6134 du CNRS
et de l'Universit\'e de Corse,\\
Universit\'e de Corse, BP 52, F-20250 Corte,
France}

\author{Mohamed {Ould El Hadj}}
\email{ould-el-hadj@univ-corse.fr}

\affiliation{Equipe Physique
Th\'eorique - Projet COMPA, \\ SPE, UMR 6134 du CNRS
et de l'Universit\'e de Corse,\\
Universit\'e de Corse, BP 52, F-20250 Corte,
France}

\date{\today}

\begin{abstract}

With the possibility of testing massive gravity in the context of black hole physics in mind, we consider the radiation produced by a particle plunging from slightly below the innermost stable circular orbit into a Schwarzschild black hole. In order to circumvent the difficulties associated with black hole perturbation theory in massive gravity, we use a toy model in which we replace the graviton field with a massive scalar field and consider a linear coupling between the particle and this field. We compute the waveform generated by the plunging particle and study its spectral content. This permits us to highlight and interpret some important effects occurring in the plunge regime which are not present for massless fields such as (i) the decreasing and vanishing, as the mass parameter increases, of the signal amplitude generated when the particle moves on quasicircular orbits near the innermost stable circular orbit and (ii) in addition to the excitation of the quasinormal modes, the excitation of the quasibound states of the black hole.

\end{abstract}


\maketitle

\section{Introduction}
\label{introduction}
In recent years, theories of massive gravity, i.e. extensions of general relativity mediated by an ultralight massive graviton, are the subject of intense activity and the internal inconsistencies present in the old Fierz-Pauli theory \cite{Fierz:1939zz,Fierz:1939ix} (in particular, the discontinuity with general relativity in the limit where the mass of the graviton is taken to zero and the ghost-problem) seem to be overcome (see, e.g., Refs.~\cite{Hinterbichler:2011tt} and \cite{deRham:2014zqa} for reviews on massive gravity and Ref.~\cite{Goldhaber:2008xy} for experimental constraints on the mass of the graviton). Because massive gravity could explain naturally the observed accelerated expansion of the present Universe, i.e. without requiring dark energy or introducing a cosmological constant (see, e.g., Sec.~12 of Ref.~\cite{deRham:2014zqa} and references therein), it has led to numerous works in cosmology. By contrast, it has rarely been the subject of works in black hole (BH) physics (see, however, Ref.~\cite{Volkov:2013roa} and references therein for recent articles dealing with BH solutions in massive gravity and Refs.~\cite{Babichev:2013una,Brito:2013wya,Brito:2013yxa} for important considerations on the problem of BH stability in massive gravity). In our opinion, it is in this particular area that, in the near future, theories of massive gravity could be directly tested with the next generations of gravitational wave detectors.

In this article, with the possibility of testing massive gravity in the context of BH physics in mind, we intend to consider the radiation produced by a ``particle'' plunging from slightly below the innermost stable circular orbit (ISCO) into a Schwarzschild BH. We shall assume an extreme mass ratio for this physical system, i.e. that the BH is much heavier than the particle. Such a hypothesis permits us to describe the emitted radiation in the framework of BH perturbations. Here, two remarks are in order:

\begin{itemize}[label={$-$}]
   \item In the context of Einstein's general relativity, the problem we consider is of fundamental importance and there exists a large literature concerning it more or less directly (see, e.g., Refs.~\cite{Detweiler:1979xr,Oohara:1984bw,Buonanno:1998gg,
Buonanno:2000ef,Ori:2000zn,Baker:2001nu,Blanchet:2005rj,Campanelli:2006gf,Damour:2007xr,Sperhake:2007gu,
Mino:2008at,Hadar:2009ip,Hadar:2011vj,Price:2013paa,Hadar:2014dpa,Hadar:2015xpa}). Indeed, the ``plunge regime'' is the last phase of the evolution of a stellar mass object orbiting near a supermassive BH (note that it can also be used to describe the late-time evolution of a binary BH) and the waveform generated during this regime encodes the final BH fingerprint.

   \item  The Schwarzschild BH, one of the most important solutions of Einstein's general relativity, is also relevant to massive gravity \cite{Volkov:2013roa}. Indeed, it is a solution of the pathology-free bimetric theory of gravity discussed by Hassan, Schmidt-May and von Strauss in Ref.~\cite{Hassan:2012wr} and obtained by extending, in curved spacetime, the fundamental work of de Rham, Gabadadze and Tolley \cite{deRham:2010ik,deRham:2010kj}.
\end{itemize}

In the context of the bimetric theory of gravity discussed in Ref.~\cite{Hassan:2012wr}, gravitational perturbations of the Schwarzschild BH have been studied by Brito, Cardoso and Pani \cite{Brito:2013wya,Brito:2013yxa}. This problem is rather difficult and leads to results which generalize the problem of massless spin-2 fluctuations in a nontrivial way. In particular, it is important to recall that (i) the Schwarzschild BH is linearly unstable for small graviton masses, (ii) the $\ell = 1$ modes (the so-called dipole modes) become dynamic and (iii) except for the odd-parity dipole modes, all the other dynamical modes are governed by two or three coupled differential equations. All these results make very difficult, in the context of massive gravity, the theoretical analysis of the gravitational radiation produced by a particle falling into a Schwarzschild BH on an arbitrary trajectory. To our knowledge, there does not exist any work dealing fully with this fundamental problem. In a recent paper \cite{Decanini:2014kha} (see also Ref.~\cite{Decanini:2014bwa} for a more complete study including analytical results and extension to other bosonic fields), we have partially considered it by only focusing on some aspects linked with the excitation of quasinormal modes (QNMs). We have shown that, in rather large domains around critical values of the graviton mass, the excitation factors of the long-lived QNMs have a strong resonant behavior which induces, as a consequence, the existence of giant and slowly decaying ringings. This unexpected effect was obtained in a rather restricted context: (i) we only focused on the odd-parity dipole modes; (ii) the source of the distortion of the BH was described by an initial value problem with Gaussian initial data; and (iii) we did not take into account the full response of the BH but only the part associated with QNMs and therefore we neglected various important effects linked with quasibound states (QBSs) which could blur the QNM contribution.

Preliminary unpublished investigations have permitted us to understand that, in massive gravity, in addition to the theoretical difficulties previously mentioned, the problem of the excitation of a Schwarzschild BH by a plunging particle is plagued by numerical instabilities. So, in order to simplify our task, both numerically and theoretically, we shall consider in this article a toy model in which we replace the massive spin-2 perturbations with a massive scalar field and consider a linear coupling between the particle and this field. We shall compute the waveform generated by the plunging particle and study its spectral content. This shall permit us to describe the excitation of the QNMs as well as that of the QBSs of the BH but also to show the influence of the mass on the amplitude of the emitted signal and, more particularly, on the part generated when the particle moves on quasicircular orbits near the ISCO. Of course, we hope to return to the same problem in the context of massive gravity but we think that the phenomena highlighted with the toy model are still present in this more physical and interesting problem. Furthermore, the lessons we have learned by working with the toy model should allow us to avoid some difficulties due to numerical instabilities.

Our paper is organized as follows. In Sec.~\ref{sec_2}, we describe our toy model, introduce our notations and establish theoretically the expression of the waveform emitted by a scalar point particle moving on an arbitrary geodesic by using the standard Green's function techniques. These results will be useful both in Sec.~\ref{sec_3} and Appendix \ref{appen_A}. In Sec.~\ref{sec_3}, from the closed-form expression of the plunge trajectory, we first obtain the general expression of the waveform generated by the plunging particle and then we focus more particularly on the $(\ell=2,m=2)$ mode of the scalar field. We study numerically the corresponding waveform (the so-called quadrupolar waveform) and its behavior as the mass parameter increases and as the observer location changes and we highlight the excitation of the QNMs and QBSs of the BH. In the Conclusion, we summarize the main results obtained in this article, we briefly deal with arbitrary $(\ell,m)$ modes of the scalar field and consider possible extensions of our work. In three appendixes, we  gather some important results which permit us to analyze and interpret the numerical results of Sec.~\ref{sec_3}. In Appendix \ref{appen_A}, we provide a simple closed form for the emitted waveform when the particle lives on the ISCO. In Appendix \ref{appen_B}, we focus on the weakest damped QNM with $\ell=2$ and construct numerically its response when it is excited by the plunging particle. In Appendix \ref{appen_C}, we consider the first QBSs with $\ell=2$ and determine their complex frequencies.

Throughout this article, we adopt units such that $\hbar = c = G = 1$ and we introduce the reduced mass parameter (a dimensionless coupling constant) ${\tilde \alpha}=2M\mu /{m_\mathrm{P}}^2$ where $M$, $\mu$ and ${m_\mathrm{P}}$ denote, respectively, the mass of the BH, the mass of the scalar field and the Planck mass.

\section{The general solution of the scalar wave equation with source}
\label{sec_2}
\subsection{Our model}
\label{sec_2_a}

We shall consider a point particle with mass $m_{0}$ and scalar charge $q$ moving along a world line $\gamma(\lambda)$ in the Schwarzschild spacetime $(\mathcal{M}, g_{\alpha\beta})$. $\lambda$ is an affine parameter and ($\mathcal{M}, g_{\alpha\beta}$) is the exterior of the Schwarzschild BH of mass $M$ defined by the metric
\begin{equation}
\label{Schw_metric}
ds^2= -\left(1-\frac{2M}{r} \right)dt^2+ \left(1-\frac{2M}{r} \right)^{-1}dr^2+ r^2 d\sigma_2^2
\end{equation}
where $d\sigma_2^2$ denotes the metric on the unit $2$-sphere $S^2$ and with the Schwarzschild coordinates $(t,r)$ which satisfy $t \in ]-\infty, +\infty[$ and $r \in ]2M,+\infty[$. In the following, we shall also use the so-called tortoise coordinate $r_\ast \in ]-\infty,+\infty[$ defined from the radial Schwarzschild coordinate $r$ by $dr/dr_\ast=(1-2M/r)$ and given by $r_\ast(r)=r+2M \ln[r/(2M)-1]$. We recall that the function $r_\ast=r_\ast(r)$
provides a bijection from $]2M,+\infty[$ to $]-\infty,+\infty[$.

The particle is coupled to a scalar field $\Phi$ with mass $\mu$ and the dynamics of the system field--particle is defined by the action (see also Ref.~\cite{Poisson:2011nh})
\begin{equation}
\label{action}
S=S_\mathrm{field}+S_\mathrm{part}+S_\mathrm{int}
\end{equation}
\noindent with
\begin{eqnarray}
\label{field_action}
S_\mathrm{field}&=&-\frac{1}{2}\int_{\mathcal{M}}d^{4}x\sqrt{-g(x)}\nonumber\\
& &\times\,\left[g^{\alpha\beta}(x)\nabla_{\alpha}\Phi(x)\nabla_{\beta}\Phi(x)+\mu^2\Phi^2(x)\right]
\end{eqnarray}
\noindent and
\begin{eqnarray}
\label{part_action}
S_\mathrm{part}&=&-m_{0}\int_{\gamma}d\tau \nonumber\\ &=&-m_{0}\int_{\gamma}\sqrt{-g_{\alpha\beta}(z(\lambda))\frac{dz^{\alpha}(\lambda)}{d\lambda}\frac{dz^{\beta}(\lambda)}{d\lambda}}d\lambda
\end{eqnarray}
\noindent and
\begin{equation}
\label{int_action}
S_\mathrm{int}=\int_{\mathcal{M}}d^{4}x\,\sqrt{-g(x)}\,\rho(x)\,\Phi(x).
\end{equation}
\noindent Here $S_\mathrm{field}=S_\mathrm{field}[\Phi,g_{\alpha\beta}]$ denotes the action of the scalar field $\Phi$. $S_\mathrm{part}=S_\mathrm{part}[z(\lambda),g_{\alpha\beta}]$ is the action of the particle, $\tau$ denotes its proper time and the equations $z^{\alpha}=z^{\alpha}(\lambda)$ describe its world line. \(S_\mathrm{int}=S_\mathrm{int}[z(\lambda),\Phi,g_{\alpha\beta}]\) is an interaction term describing the coupling of the particle and the field. It should be noted that they are coupled via the charge density
\begin{equation}
\label{source}
\rho(x)=q\int_{\gamma}d\tau\delta^{4}(x,z(\tau))=q\int_{\gamma}d\tau\,\frac{\delta^{4}(x-z(\tau))}{\sqrt{-g(x)}}
\end{equation}
associated with the charged scalar particle.

\subsection{Wave equation and Regge-Wheeler equation with source}
\label{sec_2_b}

The wave equation governing the scalar field $\Phi$ is obtained by extremization of the action (\ref{action}).
We have
\begin{equation}
\label{Wave_Equation}
(\Box\ - \mu^2)\Phi= -\rho,
\end{equation}
where the source $\rho$ is given by (\ref{source}).

Due to both the staticity and the spherical symmetry of the Schwarzschild background, we can solve Eq.~(\ref{Wave_Equation})
by introducing the decompositions
\begin{eqnarray}
\label{field_decomposition}
\Phi\left(t,r,\theta,\varphi\right)&=&\frac{1}{\sqrt{2\pi}}\int_{-\infty}^{+\infty}d\omega\, e^{-i\omega t}\nonumber\\
& & \times\sum_{\ell=0}^{+\infty}\sum_{m=-\ell}^{+\ell}\frac{\phi_{\omega\ell m}(r)}{r}Y_{\ell m}(\theta,\varphi)
\end{eqnarray}
\noindent and
\begin{eqnarray}
\label{source_decomposition}
\rho\left(t,r,\theta,\varphi\right)&=&\frac{1}{\sqrt{2\pi}}\int_{-\infty}^{+\infty}d\omega\, e^{-i\omega t}\nonumber\\
& & \times\sum_{\ell=0}^{+\infty}\sum_{m=-\ell}^{+\ell}\frac{\rho_{\omega\ell m}(r)}{r}Y_{\ell m}(\theta,\varphi)
\end{eqnarray}
where \(Y_{\ell{m}}(\theta,\varphi)\) are the standard spherical harmonics with \(\ell \in \mathbb{N}\) and \(m=-\ell,-\ell+1,\ldots,+\ell\).
The wave equation (\ref{Wave_Equation}) reduces then to the Regge-Wheeler equation with source
\begin{equation}
\label{RW_equation}
\left[\frac{d^{2}}{dr_{\ast}^{2}}+\omega^{2}-V_{\ell}(r)\right]\phi_{\omega\ell m}= -\left(1-\frac{2M}{r} \right)\rho_{\omega\ell m}.
\end{equation}
In Eq.~(\ref{RW_equation}), $V_{\ell}(r)$ denotes the Regge-Wheeler potential given by
\begin{equation}
\label{RW_potential}
V_{\ell}(r)=\left(1-\frac{2M}{r}\right)\left[\mu^2+\frac{\ell(\ell+1)}{r^{2}}+\frac{2M}{r^{3}}\right].
\end{equation}

\subsection{Construction of the Green's function}
\label{sec_2_c}

In order to solve the Regge-Wheeler equation (\ref{RW_equation}), we shall use the machinery of Green's functions (see Ref.~\cite{MorseFeshbach1953} for generalities on this topic and, e.g. Ref.~\cite{Breuer:1974uc} for its use in the context of BH physics). We consider the Green's function $G_{\omega\ell}(r_{*},{r}_{*}')$ defined by
\begin{equation}
\label{Green_Function_1}
\left[\frac{d^{2}}{dr_{\ast}^{2}}+\omega^{2}-V_{\ell}(r)\right]G_{\omega\ell}(r_{*},{r}_{*}')=-\delta(r_{*}-{r}_{*}')
\end{equation}
which can be written as
\begin{equation}
\label{Green_Function_2}
G_{\omega\ell}(r_{*},{r}_{*}')=-\frac{1}{W_{\ell}(\omega)}
\left\{
\begin{aligned}
&\!\!\phi_{\omega\ell}^{\mathrm{in}}(r_{*})\,\phi_{\omega\ell}^{\mathrm{up}}({r}_{*}'),\!\!&r_{*}<{r}_{\ast}',\\
&\!\!\phi_{\omega\ell}^{\mathrm{up}}(r_{*})\,\phi_{\omega\ell}^{\mathrm{in}}({r}_{*}'),\!\!&r_{*}>{r}_{\ast}'.
\end{aligned}
\right.
\end{equation}
\noindent Here $W_{\ell}(\omega)$ denotes the Wronskian of the functions $\phi_{\omega\ell}^{\mathrm {in}}$ and $\phi_{\omega\ell}^{\mathrm {up}}$. These two functions are linearly independent solutions of the homogenous Regge-Wheeler equation
\begin{equation}
\label{H_RW_equation}
\left[\frac{d^{2}}{dr_{\ast}^{2}}+\omega^{2}-V_{\ell}(r)\right]\phi_{\omega\ell}= 0.
\end{equation}
When \(\mathrm{Im}(\omega)>0\), $\phi_{\omega\ell}^{\mathrm {in}}$ is uniquely defined by its purely ingoing behavior at the event horizon $r=2M$ (i.e., for $r_\ast \to -\infty$)
\begin{subequations}
\label{bc_in}
\begin{equation}\label{bc_1_in}
\phi^\mathrm{in}_{\omega \ell} (r)\scriptstyle{\underset{r_\ast \to -\infty}{\sim}} \displaystyle{e^{-i\omega r_\ast}}
\end{equation}
while, at spatial infinity $r \to +\infty$ (i.e., for $r_\ast \to +\infty$), it has an
asymptotic behavior of the form
\begin{eqnarray}\label{bc_2_in}
& & \phi^\mathrm{in}_{\omega  \ell}(r) \scriptstyle{\underset{r_\ast \to +\infty}{\sim}}
\displaystyle{ \left[ \frac{\omega}{p(\omega)}
\right]^{1/2} } \nonumber \\
& & \quad \times \left(A^{(-)}_\ell (\omega) e^{-i[p(\omega)
r_\ast + [M\mu^2/p(\omega)] \ln(r/M)]}\right. \nonumber \\
& & \quad \quad  \left. + A^{(+)}_\ell (\omega) e^{+i[p(\omega) r_\ast +
[M\mu^2/p(\omega)] \ln(r/M)]} \right).
\end{eqnarray}
\end{subequations}
\noindent Similarly, $\phi^\mathrm{up}_{\omega \ell }$ is uniquely defined by its purely outgoing behavior at spatial infinity
\begin{subequations}
\label{bc_up}
\begin{equation}\label{bc_1_up}
\phi^\mathrm{up}_{\omega \ell} (r)\scriptstyle{\underset{r_\ast \to +\infty}{\sim}} \displaystyle{\left[ \frac{\omega}{p(\omega)}
\right]^{1/2}\!\!\!\!\! e^{+i[p(\omega) r_\ast + [M\mu^2/p(\omega)] \ln(r/M)]}}
\end{equation}
and, at the horizon, it has an asymptotic behavior of the form
\begin{equation}\label{bc_2_up}
\phi^\mathrm{up}_{\omega \ell }(r) \scriptstyle{\underset{r_\ast \to -\infty}{\sim}}\displaystyle{
B^{(-)}_\ell (\omega) e^{-i\omega r_\ast}  + B^{(+)}_\ell (\omega) e^{+i\omega r_\ast}}.
\end{equation}
\end{subequations}
In the previous expressions, the function $p(\omega)=\left( \omega^2 - \mu^2 \right)^{1/2}$ denotes the ``wave number''
while the coefficients $A^{(-)}_\ell (\omega)$, $A^{(+)}_\ell (\omega)$, $B^{(-)}_\ell (\omega)$ and $B^{(+)}_\ell (\omega)$ are complex amplitudes. Here, it is important to note that these coefficients as well as the in and up modes can be defined by analytic continuation on the two-sheet Riemann surface associated with $p(\omega)$. By evaluating the Wronskian $W_\ell (\omega)$ at $r_\ast \to -\infty$ and $r_\ast \to +\infty$, we obtain
\begin{equation}
\label{Well}
W_\ell (\omega) =2i\omega A^{(-)}_\ell (\omega) = 2i\omega B^{(+)}_\ell (\omega).
\end{equation}

\subsection{Solution of the wave equation with source}
\label{sec_2_d}

Using the Green's function (\ref{Green_Function_2}), we can show that the solution of the Regge-Wheeler equation with source (\ref{RW_equation}) is given by
\begin{equation}
\label{G_Sol_RW_Eq_en_ast}
\phi_{\omega\ell m}(r)=\int_{-\infty}^{+\infty}d{r}_{*}'\,G_{\omega\ell}(r_{*},{r}_{*}')
\left( 1 - \frac{2M}{r'({r}_{*}')}  \right)\rho_{\omega\ell m}({r}_{*}')
\end{equation}
or, equivalently, by
\begin{equation}
\label{G_Sol_RW_Eq}
\phi_{\omega\ell m}(r)=\int_{2M}^{+\infty} dr' \, G_{\omega\ell}(r,r')\rho_{\omega\ell m}(r').
\end{equation}
As a consequence, the response (\ref{field_decomposition}) detected by an observer at $(t,r,\theta,\varphi)$  is given by
\begin{equation}
\label{total_response}
\Phi(t,r,\theta,\varphi)=\frac{1}{r}\sum_{\ell=0}^{+\infty}\sum_{m=-\ell}^{+\ell}\phi_{\ell m}(t,r) Y_{\ell m}(\theta,\varphi)
\end{equation}
\noindent where
\begin{equation}
\label{Partial_Response_2}
\begin{split}
\phi_{\ell m}(t,r)=\frac{1}{\sqrt{2\pi}}\int_{-\infty+ic}^{+\infty+ic}d\omega\, &e^{-i \omega t}\phi_{\omega\ell m}(r)
\end{split}
\end{equation}
(here $c>0$) denotes the partial response corresponding to the $(\ell, m)$ mode.

For $r \to +\infty$, the solution (\ref{G_Sol_RW_Eq}) reduces to the asymptotic expression
\begin{eqnarray}
\label{Partial_Response_1}
\phi_{\omega\ell m}(r)&=&-\frac{1}{2i\omega A^{(-)}_\ell (\omega)}\left[\frac{\omega}{p(\omega)}\right]^{1/2}\nonumber\\
& &\times\, e^{+i[p(\omega) r_\ast + [M\mu^2/p(\omega)] \ln(r/M)]}\nonumber\\
& & \times\int_{2M}^{+\infty}dr'\,\phi_{\omega\ell}^{\mathrm{in}}(r')
\,\rho_{\omega\ell m}(r').
\end{eqnarray}
This result is a consequence of Eqs.~(\ref{Green_Function_2}), (\ref{bc_1_up}) and (\ref{Well}). However, it is very important to note that, in practice, if we consider that the observer  is located at a finite distance from the BH, it is not possible to use  the asymptotic behavior (\ref{bc_1_up}) in Eq.~(\ref{Green_Function_2}) but it is necessary to obtain $ \phi^\mathrm{up}_{\omega  \ell}(r)$ numerically by solving carefully  the Regge-Wheeler equation (\ref{H_RW_equation}): this is the price to pay in order to take into account the dispersive nature of the massive scalar field. In that case, assuming that the observer is located beyond the source, we must replace (\ref{Partial_Response_1}) with
\begin{eqnarray}
\label{Partial_Response_2_bis}
\phi_{\omega\ell m}(r)&=&-\frac{1}{2i\omega A^{(-)}_\ell (\omega)}\,\,\phi_{\omega\ell}^{\mathrm{up}}(r)\nonumber\\
& & \times\int_{2M}^{+\infty}dr'\,\phi_{\omega\ell}^{\mathrm{in}}(r')
\,\rho_{\omega\ell m}(r')
\end{eqnarray}

\noindent and then we have for the $(\ell,m)$ waveform
\begin{eqnarray}
\label{partial_response_def}
\phi_{\ell m}(t,r) &=& -\frac{1}{\sqrt{2\pi}}\int_{-\infty+ic}^{+\infty+ic}d\omega  \left(\frac{e^{- i \omega t}}{2 i \omega A_{\ell}^{(-)}(\omega)}\right)\nonumber\\
& & \times\, \phi_{\omega \ell}^{\mathrm{up}}(r)\int_{2M}^{+\infty}dr'\phi_{\omega \ell}^{\mathrm{in}}(r')\rho_{\omega \ell m}(r').
\end{eqnarray}

\subsection{General expression of the source}
\label{sec_2_e}

By extremization of the action (\ref{part_action}),  we can obtain the equations governing the timelike geodesic $\gamma$ followed by the massive particle. We shall denote by $t_{p}(\tau)$, $r_{p}(\tau)$, $\theta_{p}(\tau)$ and $\varphi_{p}(\tau)$ its coordinates and we remark that, without loss of generality, its trajectory can be considered to lie in the BH equatorial plane, so we assume that $\theta_{p}(\tau)=\pi/2$. We have
\begin{subequations}
\label{geodesic_system}
\begin{eqnarray}
& & \left(1-\frac{2M}{r_{p}}\right)\frac{dt_{p}}{d\tau}=\frac{E}{m_{0}}, \label{geodesic_1} \\
& & r_{p}^{2}\,\,\frac{d\varphi_{p}}{d\tau}=\frac{L}{m_{0}} \label{geodesic_2}
\end{eqnarray}
\noindent and
\begin{equation}
\label{geodesic_3}
\left(\frac{dr_{p}}{d\tau}\right)^{2}\!\!\!+\frac{L^{2}}{m_{0}^{2}r_{p}^{2}}\left(1-\frac{2M}{r_{p}}\right)-\frac{2M}{r_{p}}
=\left[\!\!\left(\frac{E}{m_{0}}\right)^{2}\!\!\!\!-1\right]\!\!,
\end{equation}
\end{subequations}
\noindent where $E$ and $L$ are, respectively, the energy and the angular momentum of the particle. Here, it is important to recall that $E$ and $L$ are two conserved quantities.

For the general trajectory previously described, the expression (\ref{source}) of the source is then given by
\begin{eqnarray}
\label{Density_Source_1}
\rho(x)= q\,\delta(\theta-&&\pi/2)\int_{\gamma}\frac{d\tau}{r^{2}_{p}(\tau)}\,\delta(t-t_{p}(\tau))\nonumber\\
\times&&\,\delta(r-r_{p}(\tau))\,\delta(\varphi-\varphi_{p}(\tau)).
\end{eqnarray}

\section{Waveforms emitted by a scalar point particle on a plunge trajectory}
\label{sec_3}

\subsection{Source due to a scalar point particle on a plunge trajectory}
\label{sec_3_a}

In this section, we consider a particle plunging into the BH from the ISCO at $r_{\mathrm{ISCO}}= 6M$ or, more precisely, from slightly below the ISCO. In order to describe its trajectory, we need its angular momentum and its energy on the ISCO. They are determined by using  Eqs.~(\ref{angular_momentum_circular}) and (\ref{energy_circular}) which have been established for an arbitrary circular trajectory. We have
\begin{subequations}
\begin{equation}
\frac{L_{\mathrm{ISCO}}}{m_{0}}=2\sqrt{3}M
\end{equation}
\noindent and
\begin{equation}
\frac{E_{\mathrm{ISCO}}}{m_{0}}=\frac{2\sqrt{2}}{3}.
\end{equation}
\end{subequations}

Substituting $L_{\mathrm{ISCO}}$ and $E_{\mathrm{ISCO}}$ into the geodesic equations (\ref{geodesic_1})-(\ref{geodesic_3}), we obtain after integration

\begin{eqnarray}
\label{trajectory_plung}
\frac{t_{p}(r)}{2M}\!&=&\!\frac{2\sqrt{2}\left(r-24M\right)}{2M\left(6M/r-1\right)^{1/2}}-22\sqrt{2}
\tan^{-1}\!\!\left[\left(6M/r-1\right)^{1/2}\right]\nonumber\\
&&+ 2\tanh^{-1}\!\!\left[\left(3M/r-1/2\right)^{1/2}\right]+  \frac{t_{0}}{2M}
\end{eqnarray}
\noindent and
\begin{equation}
\label{trajectory_plung_phi}
\varphi_{p}(r)=-\frac{2\sqrt{3r}}{\left(6M-r\right)^{1/2}}+\varphi_{0}
\end{equation}

\noindent where $t_{0}$ and $\varphi_{0}$ are two arbitrary integration constants. From (\ref{trajectory_plung_phi}), we can write the spatial trajectory of the plunging particle in the form
\begin{equation}
\label{trajectory_plung_phi_bis}
r_{p}(\varphi)=\frac{6M}{[1+12/(\varphi-\varphi_{0})^{2}]}.
\end{equation}

\begin{figure}[h!]
\centering
\includegraphics[scale=0.6]{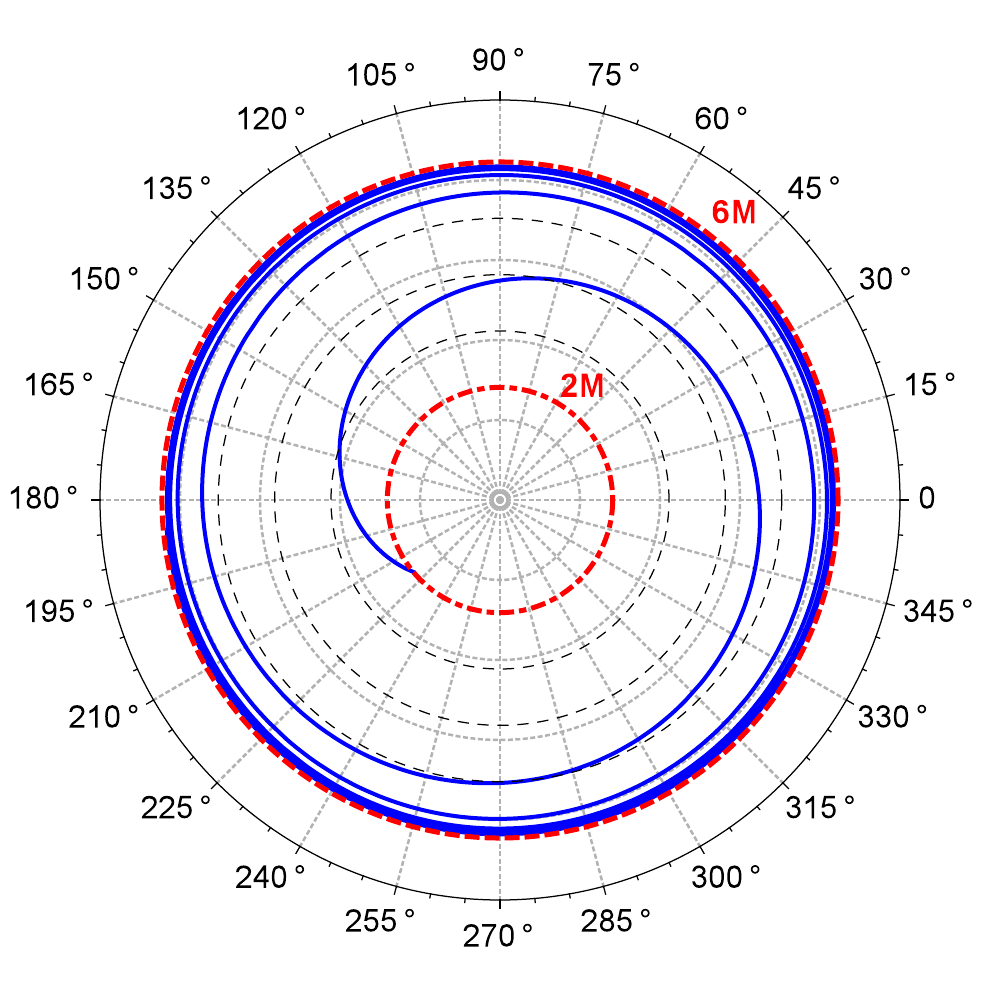}
\setlength\abovecaptionskip{-0.5ex}
\caption{\label{Trajectory_Plung} The plunge trajectory obtained from Eq.~(\ref{trajectory_plung_phi_bis}). Here, we assume that the particle starts at $r=r_{\mathrm{ISCO}}(1-\epsilon)$ with $\epsilon=10^{-2}$ and we take $\varphi_{0}=0$. The red dashed line at $r=6M$ and the red dot-dashed line at $r=2M$ represent the ISCO and the horizon, respectively.}
\end{figure}

\noindent This trajectory is displayed in Fig.~\ref{Trajectory_Plung}.

After integration and by using, in particular, the change of variable $\tau\rightarrow \varphi_{p}(\tau)$, Eq.~(\ref{Density_Source_1}) leads to

\begin{eqnarray}
\label{Density_Source_plung_1}
\rho(t,r,\theta&,&\varphi)=\frac{1}{2\pi}\int_{-\infty}^{+\infty}\!\!\!\!d\omega\, e^{- i \omega t}\sum_{\ell=0}^{+\infty}\sum_{m=-\ell}^{+\ell}
\left\{\!\vphantom{\frac {e^{i \omega t_{p}(r)}}{\left(3\,r_{s}-r\right)^{3/2}}} \frac{3\,q}{\sqrt{r}}\right.\nonumber\\
&&\left.\frac {e^{i \left[\omega t_{p}\left(r\right)- m\varphi_{p}(r)\right]}}{\left(6M-r\right)^{3/2}} Y_{\ell m}^{*}\left(\frac{\pi}{2},0\right)\vphantom{\left(1-\frac{3M}{r_{0}}\right)^{1/2}}\!\!\!\right\}Y_{\ell m}(\theta,\phi)
\end{eqnarray}

\noindent and we can then show  that (\ref{Density_Source_plung_1}) can be written in the form (\ref{source_decomposition}) with

\begin{equation}
\label{term_source_plung}
\rho_{\omega \ell m}(r)=\frac{3\,q\sqrt{r}}{\sqrt{2\pi}}\,
\frac {e^{i \left[\omega t_{p}(r)- m\varphi_{p}(r)\right]}}{\left(6M-r\right)^{3/2}}\, Y_{\ell m}^{*}\left(\frac{\pi}{2},0\right).
\end{equation}

\subsection{Quadrupolar waveform produced by the plunging particle}

\label{sec_3_b}

\begin{figure}[h]
\centering
       \includegraphics[scale=0.5]{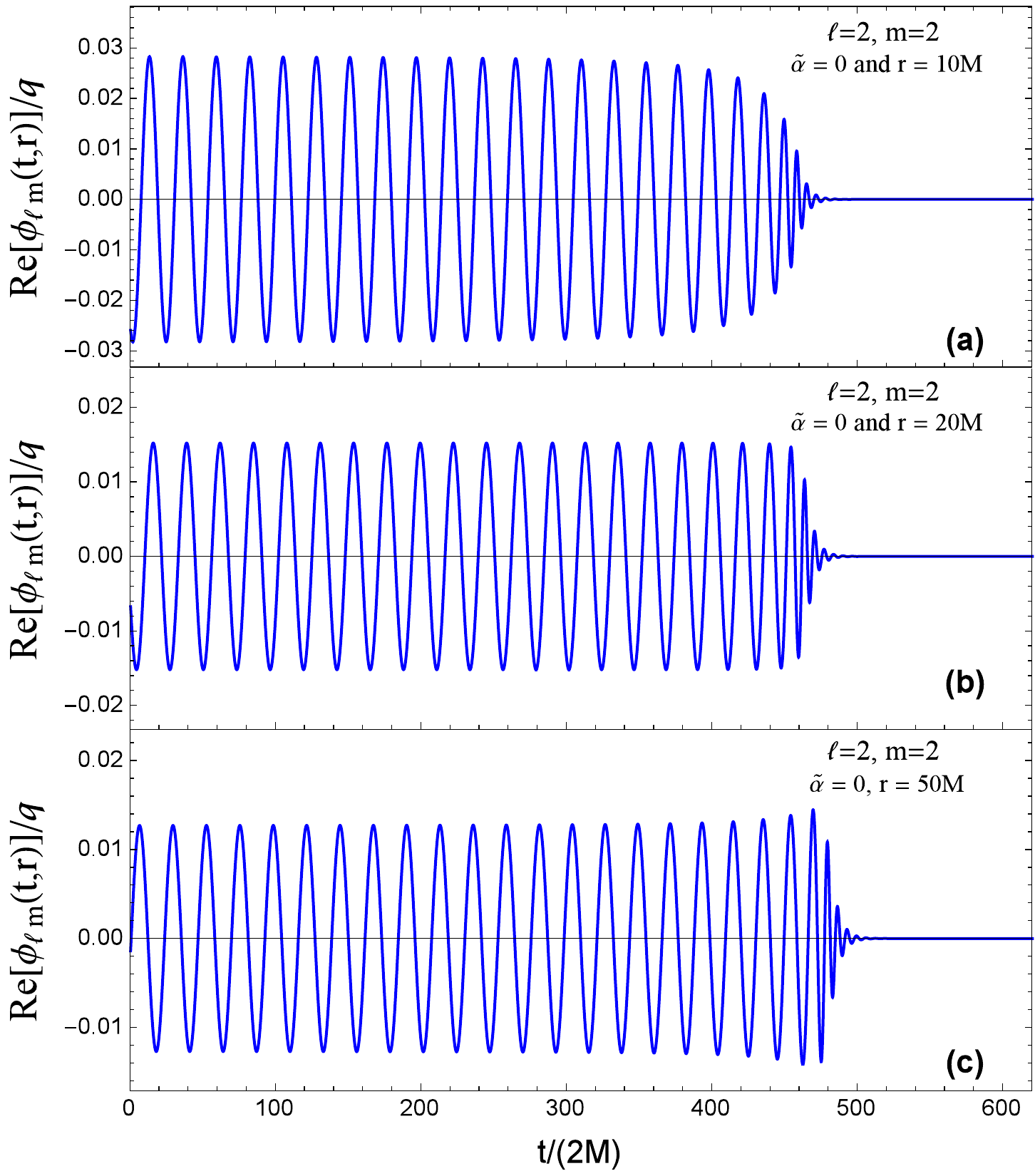}
        \vspace{-0.3cm}
\caption{\label{Rep_Plug_mu_0} Quadrupolar waveforms produced by the plunging particle. The results are obtained for a massless scalar field  $(\tilde\alpha=0)$ and the observer is located at (a) $r=10M$, (b) $r=20M$ and (c) $r=50M$.}
\end{figure}

\begin{figure}[h]
\centering
       \includegraphics[scale=0.5]{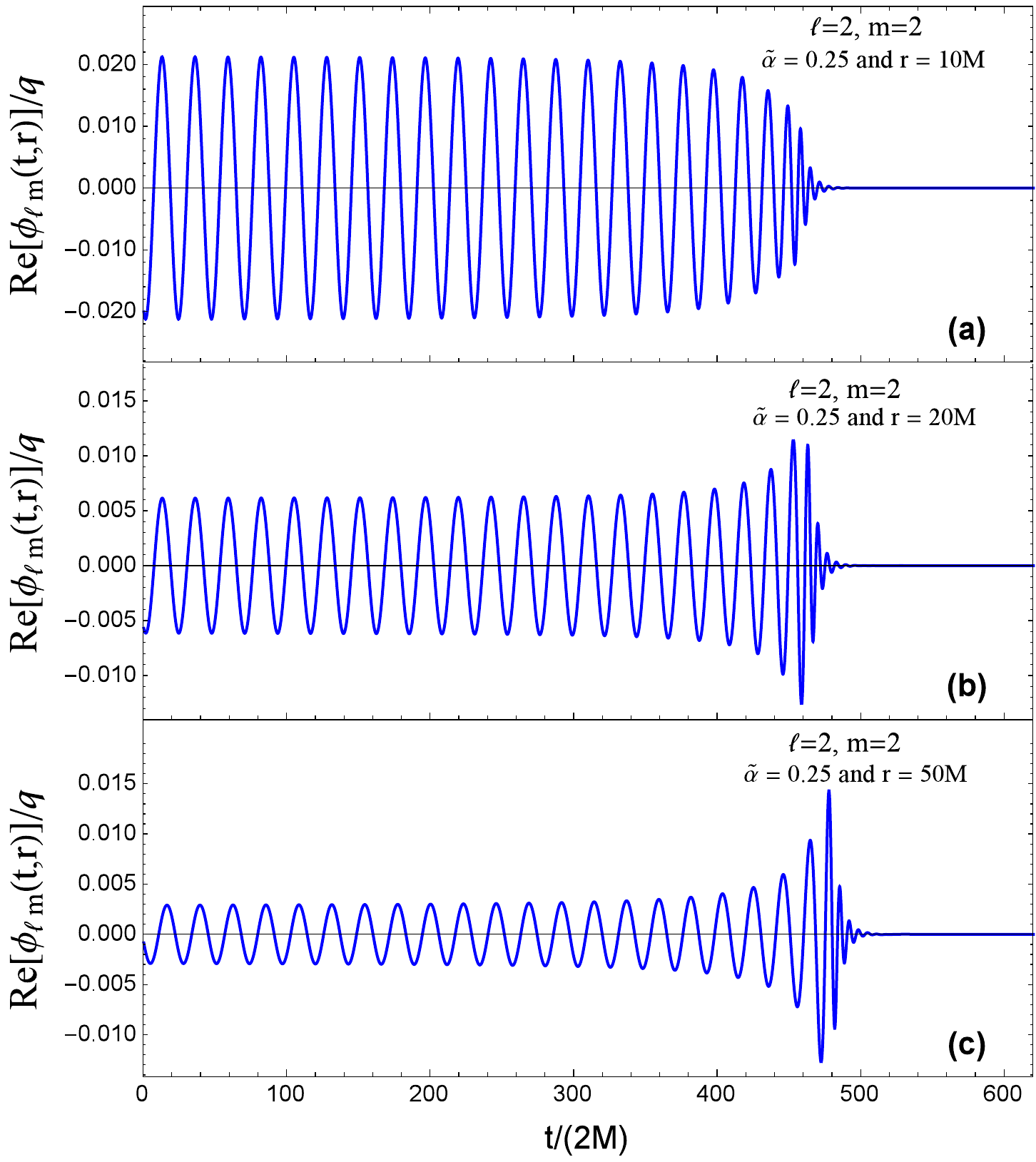}
        \vspace{-0.3cm}
\caption{\label{Rep_Plug_mu_025} Quadrupolar waveforms produced by the plunging particle. The results are obtained for a massive scalar field  ($\tilde\alpha=0.25$ is below the threshold $\tilde\alpha_c \approx0.2722 $) and the observer is located at (a) $r=10M$, (b) $r=20M$ and (c) $r=50M$.}
\end{figure}

\begin{figure}[h!]
\centering
       \includegraphics[scale=0.5]{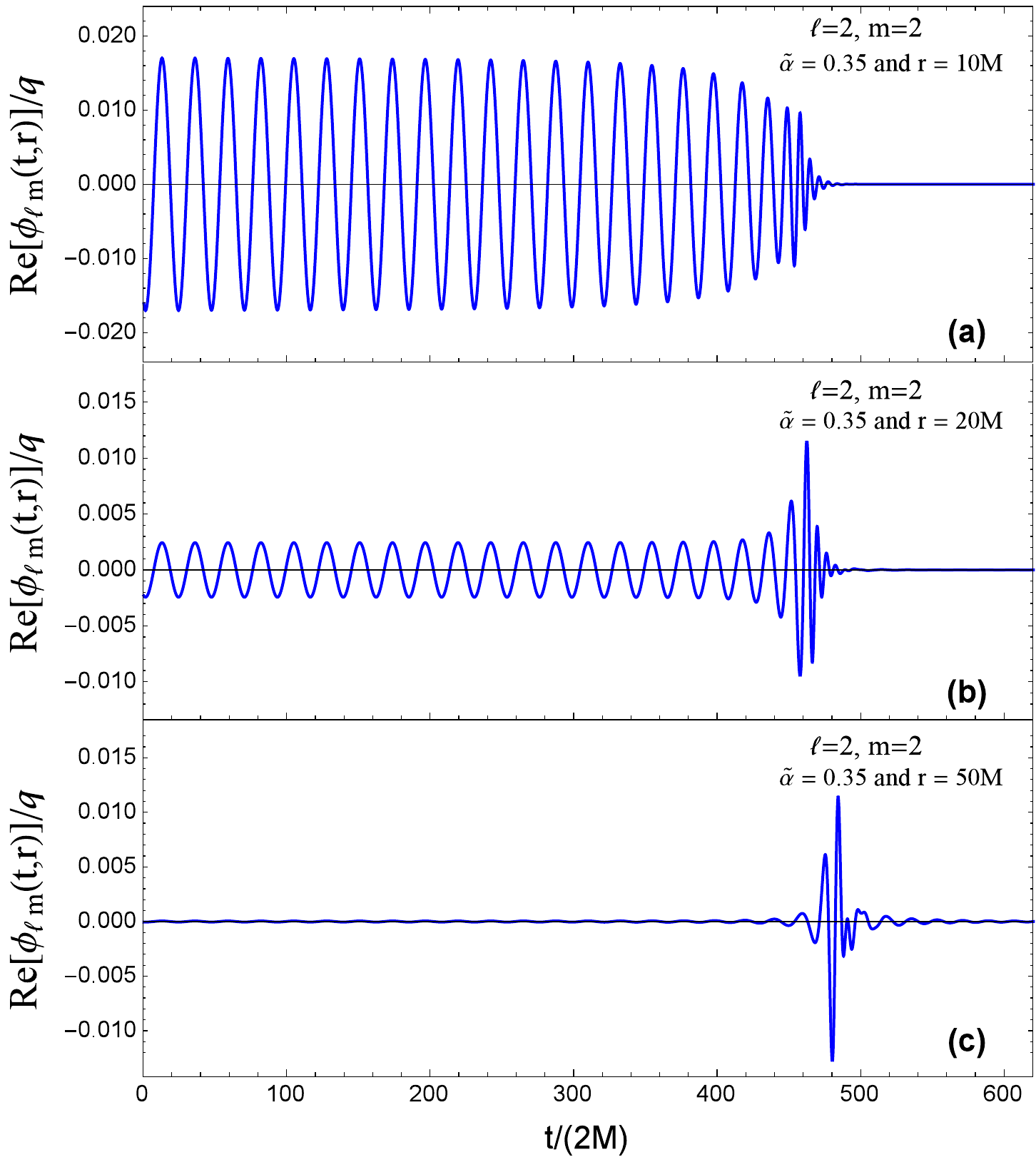}
          \vspace{-.3cm}
\caption{\label{Rep_Plug_mu_035} Quadrupolar waveforms produced by the plunging particle. The results are obtained for a massive scalar field  ($\tilde\alpha=0.35$ is above the threshold $\tilde\alpha_c \approx0.2722 $)and the observer is located at (a) $r=10M$, (b) $r=20M$ and (c) $r=50M$.}
\end{figure}

From Eq.~(\ref{partial_response_def}) with the source term given by Eq.~(\ref{term_source_plung}), we can now obtain numerically the partial waveform $\phi_{\ell m}(t,r)$ emitted by the plunging particle. In this section, we shall only focus on the $(\ell=2, m=2)$ mode of the scalar field and we shall emphasize the role of the mass parameter $\mu$ and the location of the observer.

In Figs.~\ref{Rep_Plug_mu_0}, \ref{Rep_Plug_mu_025} and \ref{Rep_Plug_mu_035}, we display the partial waveforms corresponding, respectively, to the values ${\tilde \alpha}=0$, $0.25$ and $0.35$ of the reduced mass parameter. For each one, we consider that the observer is located at $r=10M$, $20M$ and $50M$. The waveforms have been obtained by assuming that the particle starts at $r=r_\mathrm{ISCO}(1-\epsilon)$ with $\epsilon=10^{-4}$ and, furthermore, we have taken  $\varphi_{0}=0$ and $t_{0}/(2M)=500$ in order to shift the interesting part of the signal in the window $t/(2M)\in[0,600]$. To construct these waveforms and, in particular, to obtain the functions  $\phi^\mathrm{in}_{\omega \ell}$ and $\phi^\mathrm{up}_{\omega \ell}$ as well as the coefficient  $A^{(-)}_\ell (\omega)$, we have numerically integrated the Regge-Wheeler equation (\ref{H_RW_equation}) with the Runge-Kutta method. The initialization of the process has been achieved with Taylor series expansions converging near the horizon and we have compared the solutions to asymptotic expansions with ingoing and outgoing at spatial infinity that we have decoded by Pad\'e summation. Moreover, in Eq.~(\ref{partial_response_def}), we have discretized the integral over $\omega$. For ${\tilde \alpha}=0$, $0.25$ and $0.35$, in order to obtain numerically stable waveforms, we can limit the range of frequencies to $-6\leq 2M\omega \leq+6$ and take for the frequency resolution $2M\delta\omega=1/10000$. It should be noted, however, that in order to extract correctly the spectral content of the signals (see Sec.~\ref{sec_3_e}), it is necessary to work with higher-frequency resolutions which strongly depend on the mass parameter ${\tilde \alpha}$. For example, for ${\tilde \alpha}=0$ it is sufficient to take $2M\delta\omega=1/10000$, while for ${\tilde \alpha}=0.35$, we have worked with $2M\delta\omega=1/100000$.

For the massless scalar field (see Fig.~\ref{Rep_Plug_mu_0}), the waveforms can be decomposed in three phases: (i) an ``adiabatic phase'' corresponding to the quasicircular motion of the particle near the ISCO (see Fig.~\ref{Trajectory_Plung}), (ii) a ringdown phase due to the excitation of QNMs and (iii) a late-time phase. Such a decomposition remains roughly valid for the massive field (compare Figs.~\ref{Rep_Plug_mu_025} and \ref{Rep_Plug_mu_035} with Fig.~\ref{Rep_Plug_mu_0}) but now, as we shall see in more detail in the following subsections, the behavior of the signal is modified by the excitation of the QBSs and, moreover, at a large distance from the BH, it strongly depends on a threshold value $\tilde\alpha_{c}$ of the dimensionless coupling constant $\tilde\alpha$ which separates the regions where the $(\ell,m)$ mode of the field is propagating (but dispersive) or evanescent. This threshold corresponds to the mass parameter $\mu_c = 2 \Omega_{\mathrm{ISCO}}$ where $\Omega_{\mathrm{ISCO}}$ denotes the angular velocity of the particle moving on the ISCO or, in other words, its orbital frequency. The existence of this threshold and its consequences are discussed in detail in Appendix A.

\subsection{The adiabatic phase and the circular motion of the particle on the ISCO}
\label{sec_3_c}

\begin{figure}[h!]
\centering
          \includegraphics[scale=0.5]{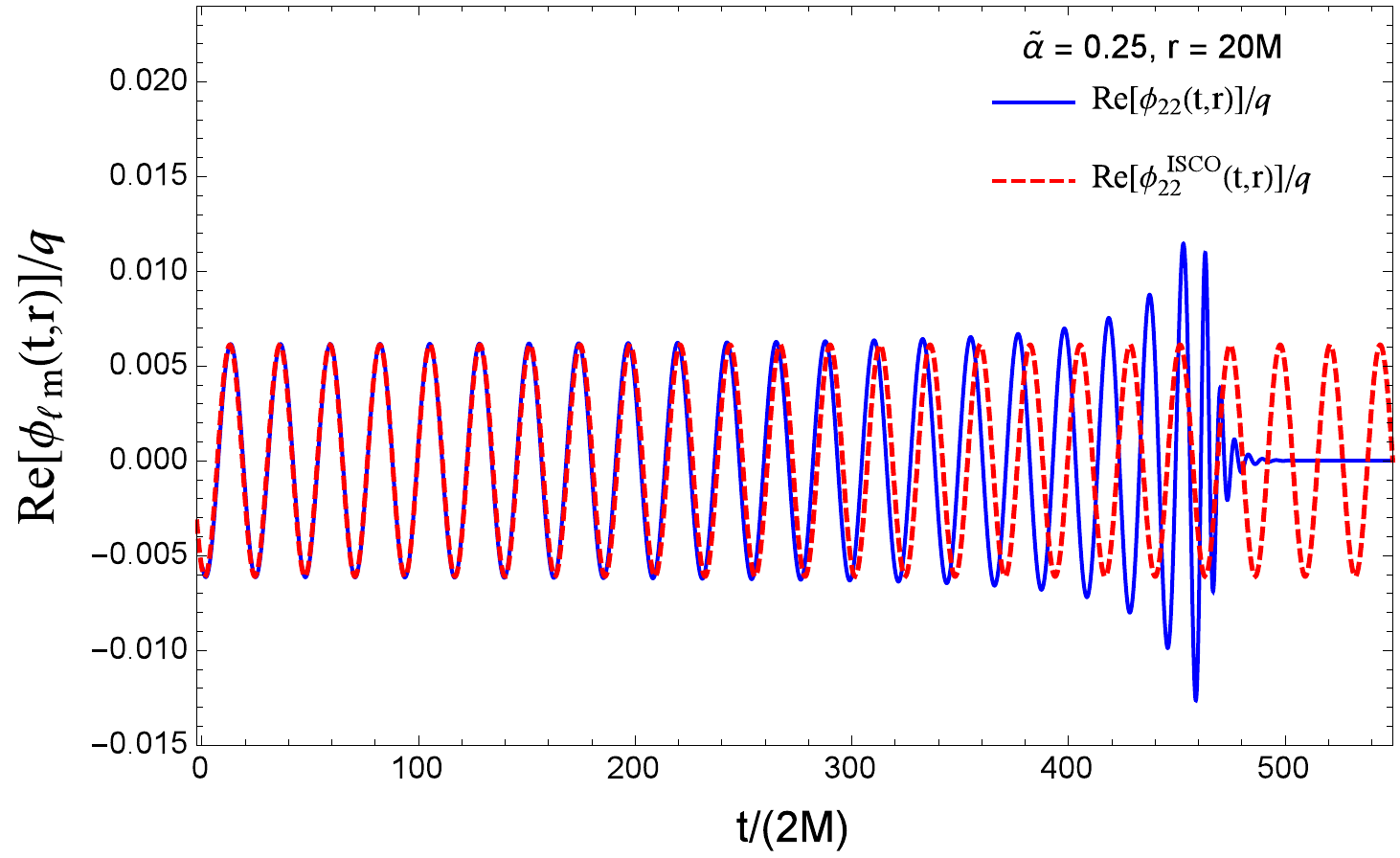}
\caption{\label{Circular_Plung_2_2_mu_025_20M} Quadrupolar waveforms produced by the plunging particle (blue line) and by a particle orbiting the BH on the ISCO (red dashed line). The reduced mass $(\tilde\alpha=0.25)$ is below the threshold $(\tilde\alpha_{c}\approx0.2722)$.}
\end{figure}

\begin{figure}[h]
\centering
      \includegraphics[scale=0.5]{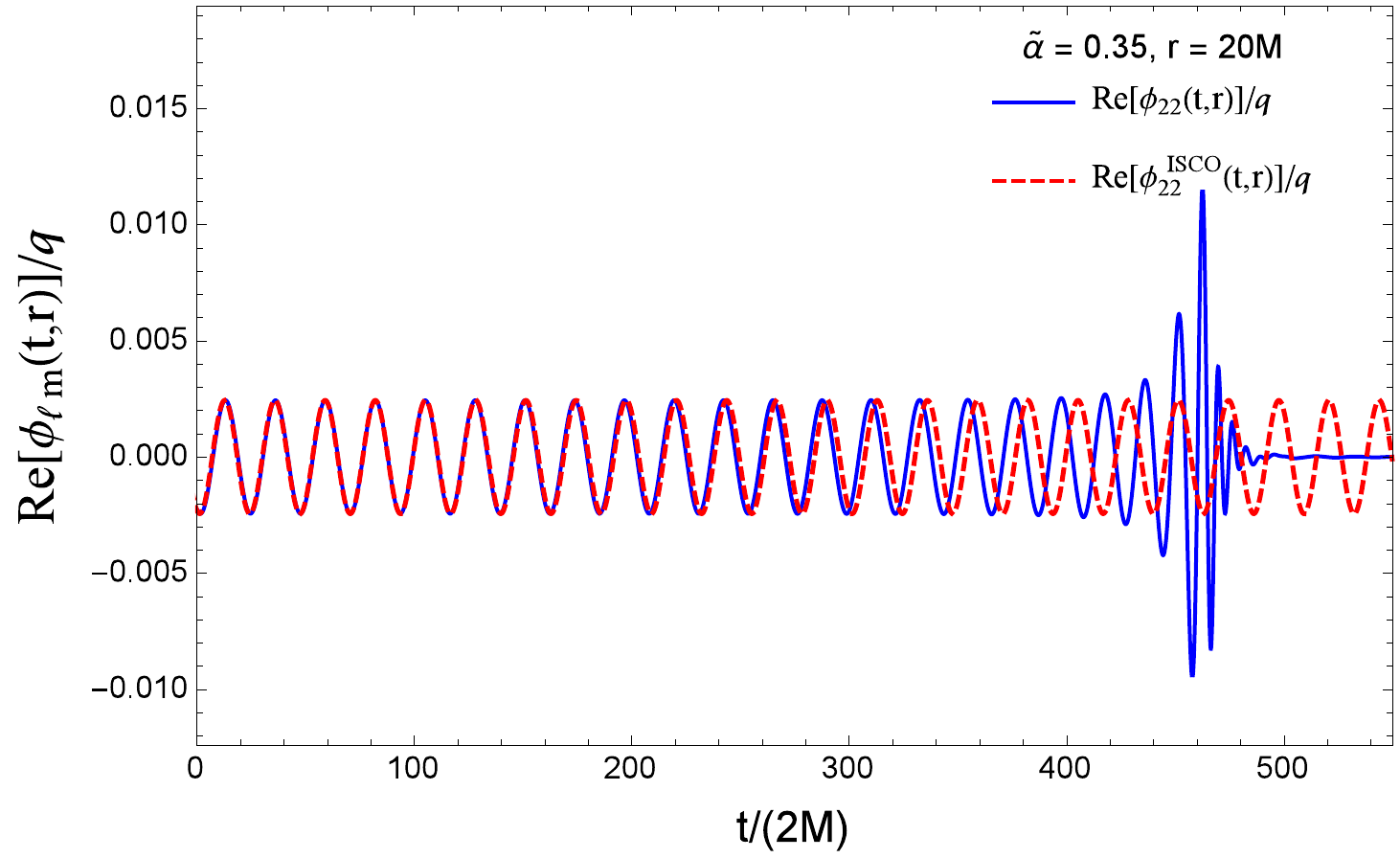}
       \vspace{-.3cm}
\caption{\label{Circular_Plung_2_2_mu_035_20M} Quadrupolar waveforms produced by the plunging particle (blue line) and by a particle orbiting the BH on the ISCO (red dashed line). The reduced mass $(\tilde\alpha=0.35)$ is above the threshold $(\tilde\alpha_{c}\approx0.2722)$.}
\end{figure}

In Figs.~\ref{Circular_Plung_2_2_mu_025_20M} and \ref{Circular_Plung_2_2_mu_035_20M}, we compare the quadrupolar waveform produced by the plunging particle we have obtained in Sec.~\ref{sec_3_b} with the quadrupolar waveform produced by a particle orbiting the BH on the ISCO we discuss in Appendix \ref{appen_A}. We consider two particular masses given by $\tilde\alpha=0.25$ and $\tilde\alpha=0.35$, i.e. one below the threshold value $\tilde\alpha_{c}\approx0.2722$ and the other one above, but the discussion remains valid even for other masses if they are not too large. In the adiabatic phase, the waveform emitted by the plunging particle is described very accurately by the waveform emitted by the particle living on the ISCO. This can be easily  understood by noting that the initial position of the plunging particle is very close to the ISCO so that it undergoes an adiabatic inspiral along a sequence of quasicircular orbits near the ISCO. As a consequence, the comments made in Appendix \ref{appen_A} (see also Fig.~\ref{Amp_Rep_Circ_mu_l_2_10M_20M_200M}) permit us to understand the behavior of the waveform in the adiabatic phase and to interpret Figs.~\ref{Rep_Plug_mu_025} and \ref{Rep_Plug_mu_035}. It is, in particular, important to note that in the adiabatic phase:
\begin{itemize}[label={$-$}]
  \item For a given ``distance'' $r$ of the observer, the waveform amplitude decreases as the mass increases and vanishes for large masses.
  \item Above the threshold, for an observer at spatial infinity, the waveform amplitude vanishes. (However, as  we shall see in Sec.~\ref{sec_3_e}, this result is slightly modified due to the excitation of the long-lived QBSs).
\end{itemize}

\subsection{The ringdown phase and the excitation of QNMs}
\label{sec_3_d}

\begin{figure}[h]
\centering
      \includegraphics[scale=0.5]{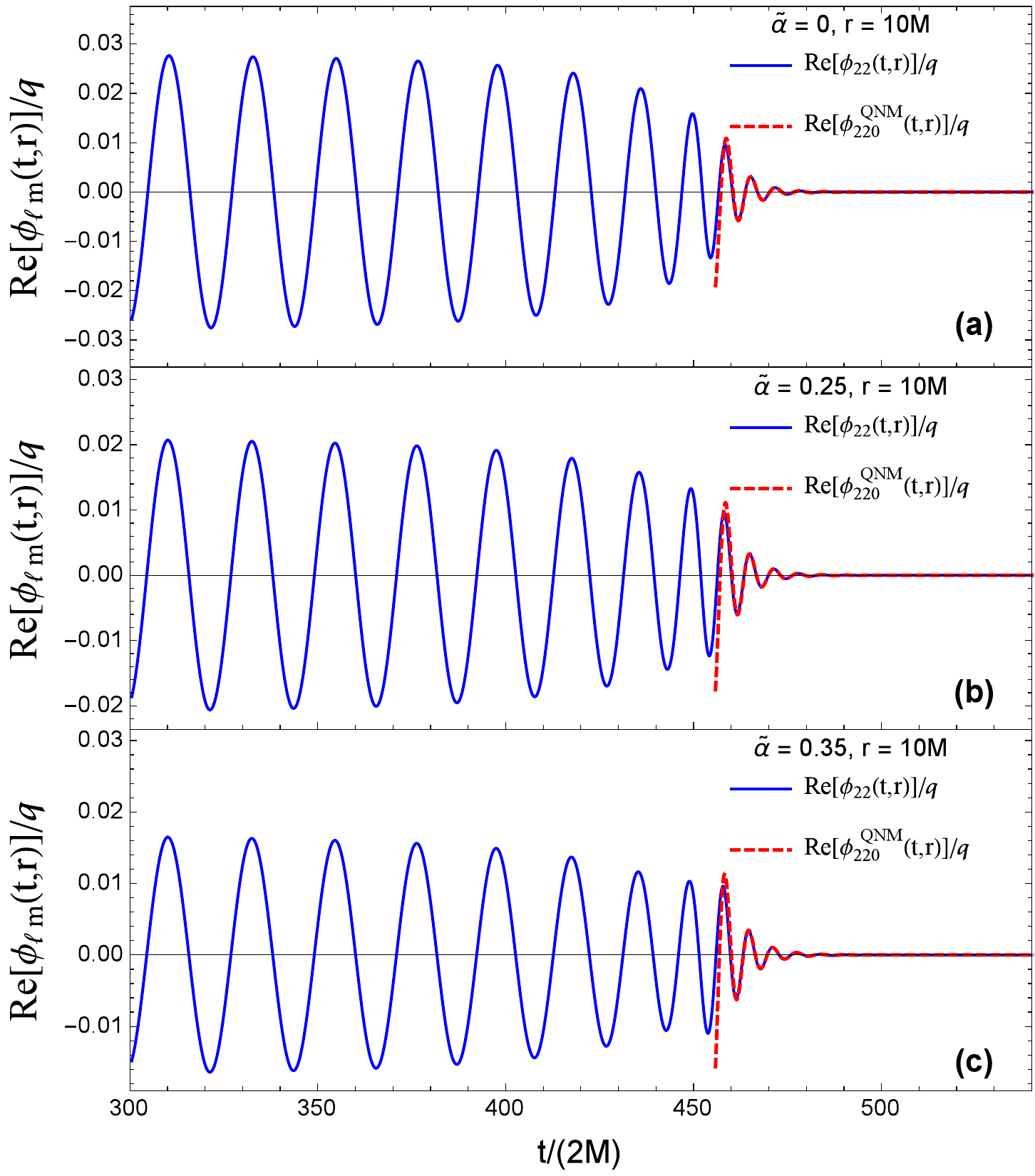}
        \vspace{-.3cm}
\caption{\label{QNM_2_0_mu_0_10M} Comparison of the quadrupolar waveform produced by the plunging particle (blue line) and the quadrupolar quasinormal waveform (red dashed line). The results are obtained for an observer at $r=10M$.}
\end{figure}

\begin{figure}[h]
\centering
      \includegraphics[scale=0.5]{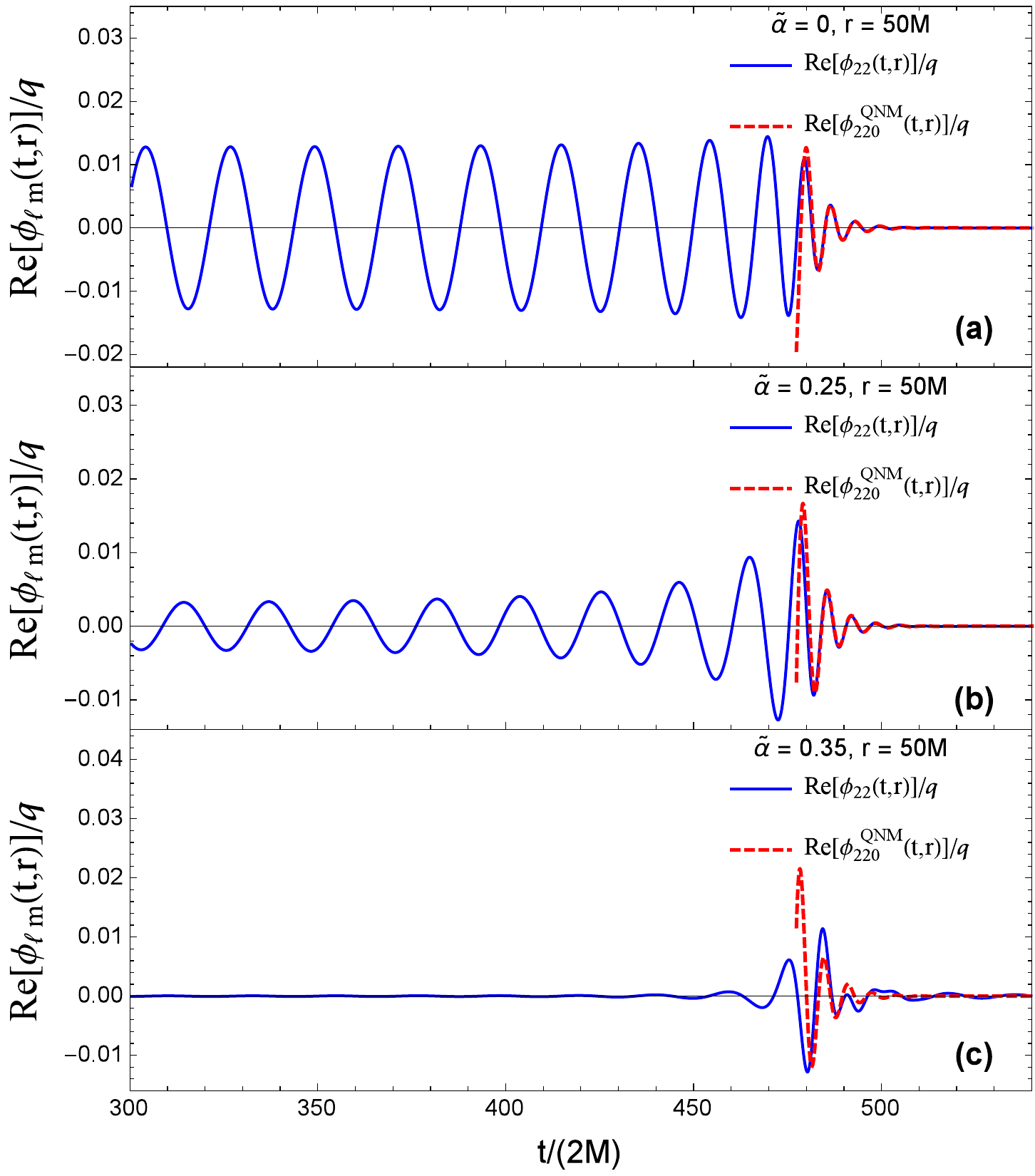}
        \vspace{-.3cm}
\caption{\label{QNM_2_0_mu_0_50M} Comparison of the quadrupolar waveform produced by the plunging particle (blue line) and the quadrupolar quasinormal waveform (red dashed line). The results are obtained for an observer at $r=50M$.}
\end{figure}

In Figs.~\ref{QNM_2_0_mu_0_10M} and \ref{QNM_2_0_mu_0_50M}, we compare the quadrupolar waveform produced by the plunging particle we have obtained in Sec.~\ref{sec_3_b} with the quadrupolar quasinormal waveform $\phi_{220}^{\mathrm{QNM}}$ given by Eq.~(\ref{phi_QNM_2}) which corresponds to the fundamental ($\ell=2, n=0$) QNM
we discuss in Appendix \ref{appen_B}. We have considered two locations for the observer ($r=10M$ and $r=50M$) and three particular reduced masses given by $\tilde\alpha=0$, $0.25$ and $0.35$. When the reduced mass $\tilde\alpha$ and the distance $r$ are not too large, the quasinormal waveform describes very accurately the ringdown phase. However, we have checked that if the reduced mass $\tilde\alpha$ or the distance $r$  increases, the agreement is not so good (cf., e.g. Fig~\ref{QNM_2_0_mu_0_50M}c where $\tilde\alpha$ and $r$ are both large). In our opinion, this is due to the excitation of QBSs (see Sec.~\ref{sec_3_e}) which blur the QNM contribution.

\subsection{Excitation of QBSs}
\label{sec_3_e}

In this subsection, we focus on the spectral content of the waveform by considering separately the adiabatic and the late-time phases. The associated spectral contents can be obtained by using the Fourier transform machinery and limiting the time integrations to the considered phase.

\begin{figure}[h]
\centering
       \includegraphics[scale=0.5]{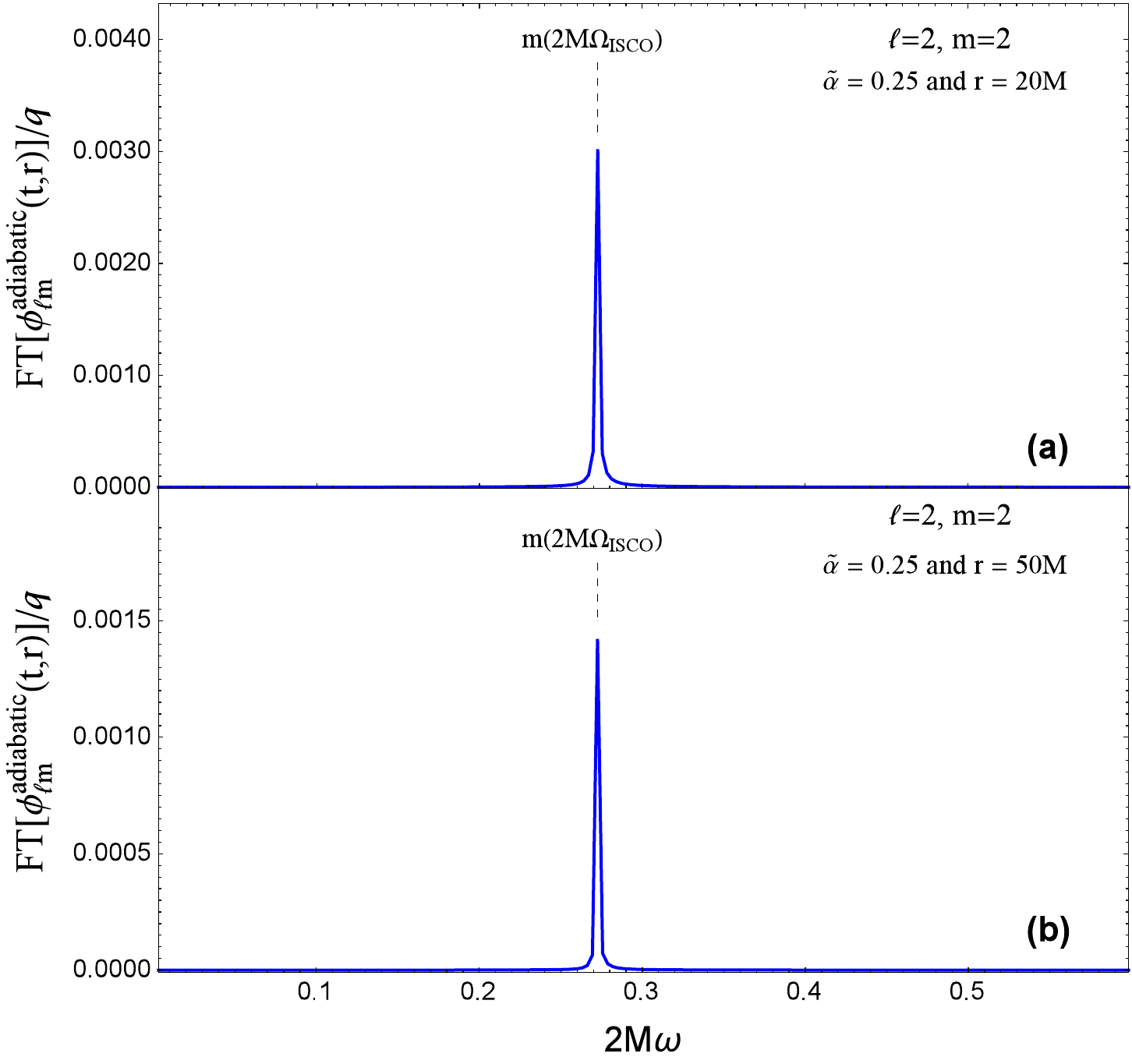}
          \vspace{-.3cm}
\caption{\label{Angular_velocity_mu_025} Spectral content of the adiabatic phase of the quadrupolar waveform produced by the plunging particle. The results are obtained for a massive scalar field  ($\tilde\alpha=0.25$ is below the threshold $\tilde\alpha_c \approx0.2722 $) and the observer is located at (a) $r=20M$ and (b) $r=50M$. We observe the signature of the quasicircular motion of the plunging particle.}
\end{figure}

\begin{figure}[h]
\centering
       \includegraphics[scale=0.5]{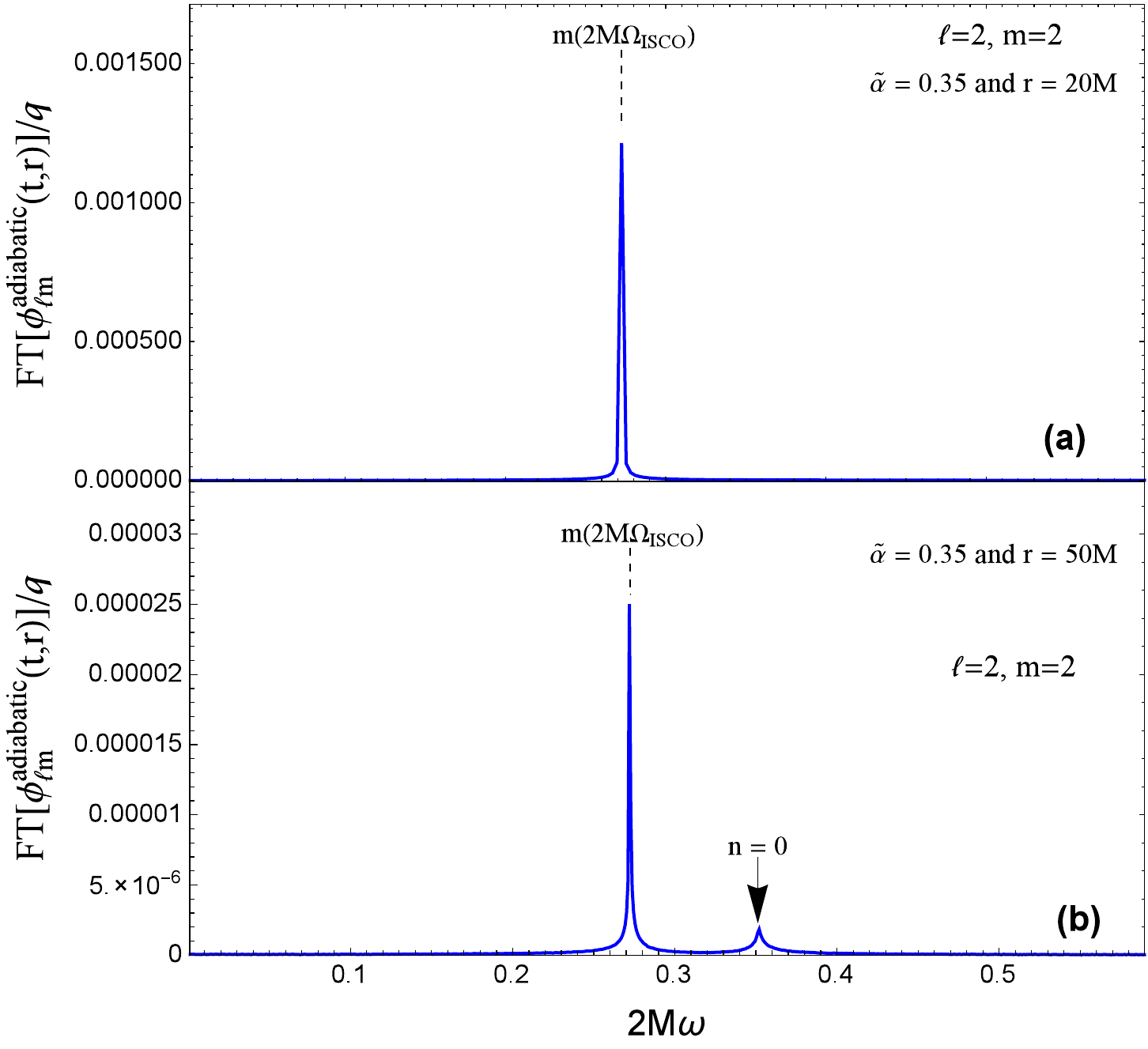}
          \vspace{-.3cm}
\caption{\label{Angular_velocity_mu_035} Spectral content of the adiabatic phase of the quadrupolar waveform produced by the plunging particle. The results are obtained for a massive scalar field  ($\tilde\alpha=0.35$ is above the threshold $\tilde\alpha_c \approx0.2722 $) and the observer is located at (a) $r=20M$ and (b) $r=50M$. We observe, in addition to the signature of the quasicircular motion of the plunging particle, that of the first long-lived QBS.}
\end{figure}

In Figs.~\ref{Angular_velocity_mu_025} and \ref{Angular_velocity_mu_035},  we display the spectral content of the adiabatic phase of the quadrupolar waveform produced by the plunging particle. For a reduced mass parameter $\tilde\alpha$ below the threshold $\tilde\alpha_{c}$ (see Fig.~\ref{Angular_velocity_mu_025}), we can observe a peak at $\omega=2\Omega_{ISCO}$. It is, of course, associated with the quasicircular motion of the plunging particle near the ISCO.  For a reduced mass parameter $\tilde\alpha$ above the threshold $\tilde\alpha_{c}$ (see Fig.~\ref{Angular_velocity_mu_035}), the same peak at $\omega=2\Omega_{ISCO}$ is present. However, it is important to note that its height decreases very rapidly as the distance $r$ of the observer increases. This is due to the evanescent character of the ($\ell=2, m=2$) mode below $\tilde\alpha_{c}$. It is, moreover, very interesting to remark the presence of another peak at a frequency equals to the real part of the complex frequency of the first long-lived QBS (see Table~\ref{tab:QBS}). In other terms, we can observe the excitation of the first QBS in the adiabatic phase! As we have already noted in Sec.~\ref{sec_3_c}, above the threshold, for an observer at a large distance from the BH, the contribution of the quadrupolar waveform associated with the quasicircular motion vanishes but the QBSs pick up the slack.

\begin{figure}[h]
\centering
      \includegraphics[scale=0.5]{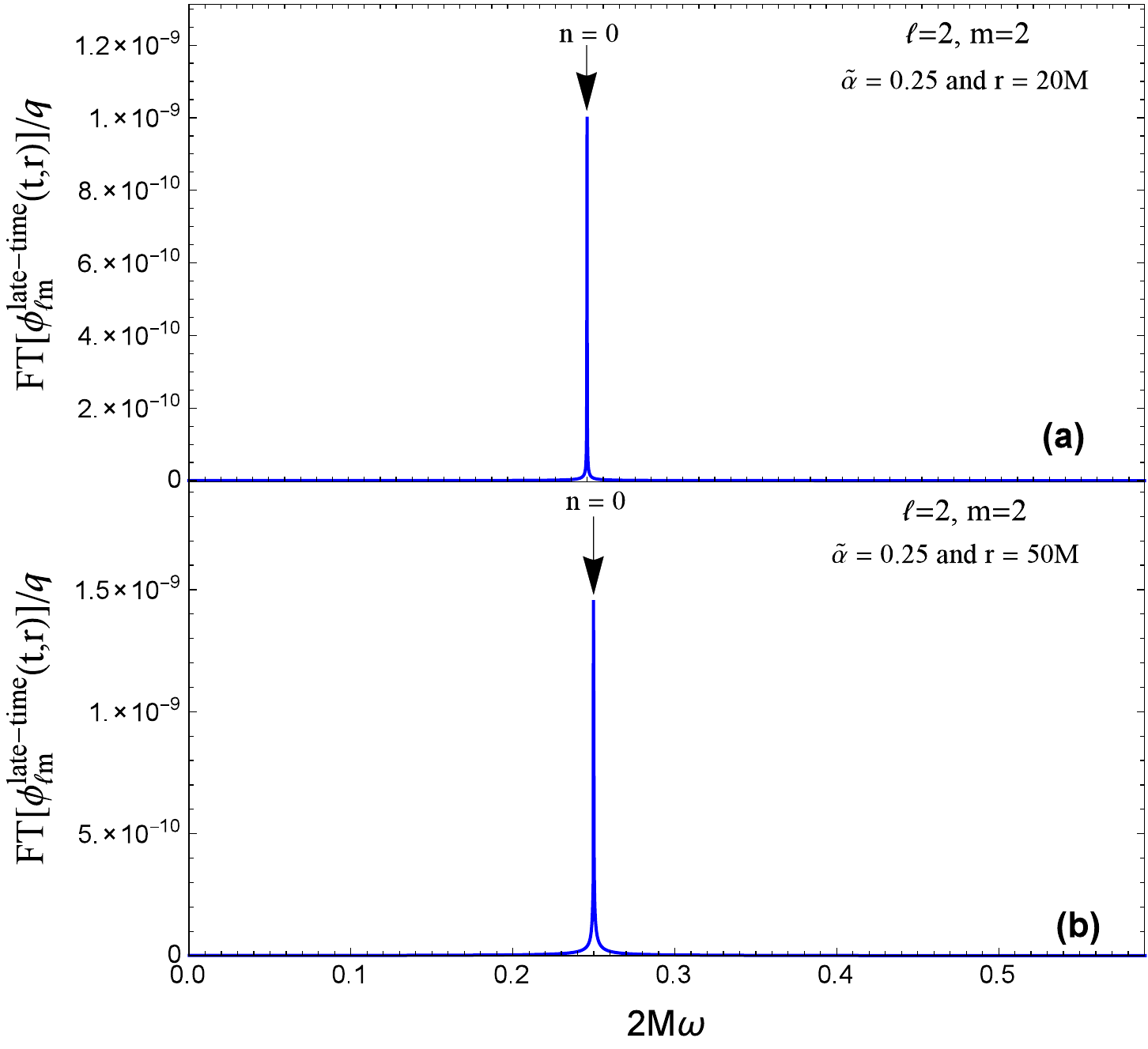}
           \vspace{-.3cm}
\caption{\label{QBS_2_0_mu_025} Spectral content of the late-time phase of the quadrupolar waveform produced by the plunging particle. The results are obtained for a massive scalar field  ($\tilde\alpha=0.25$ is below the threshold $\tilde\alpha_c \approx0.2722 $) and the observer is located at (a) $r=20M$ and (b) $r=50M$. We observe the signature of the first long-lived QBS.}
\end{figure}

\begin{figure}[h]
\centering
       \includegraphics[scale=0.5]{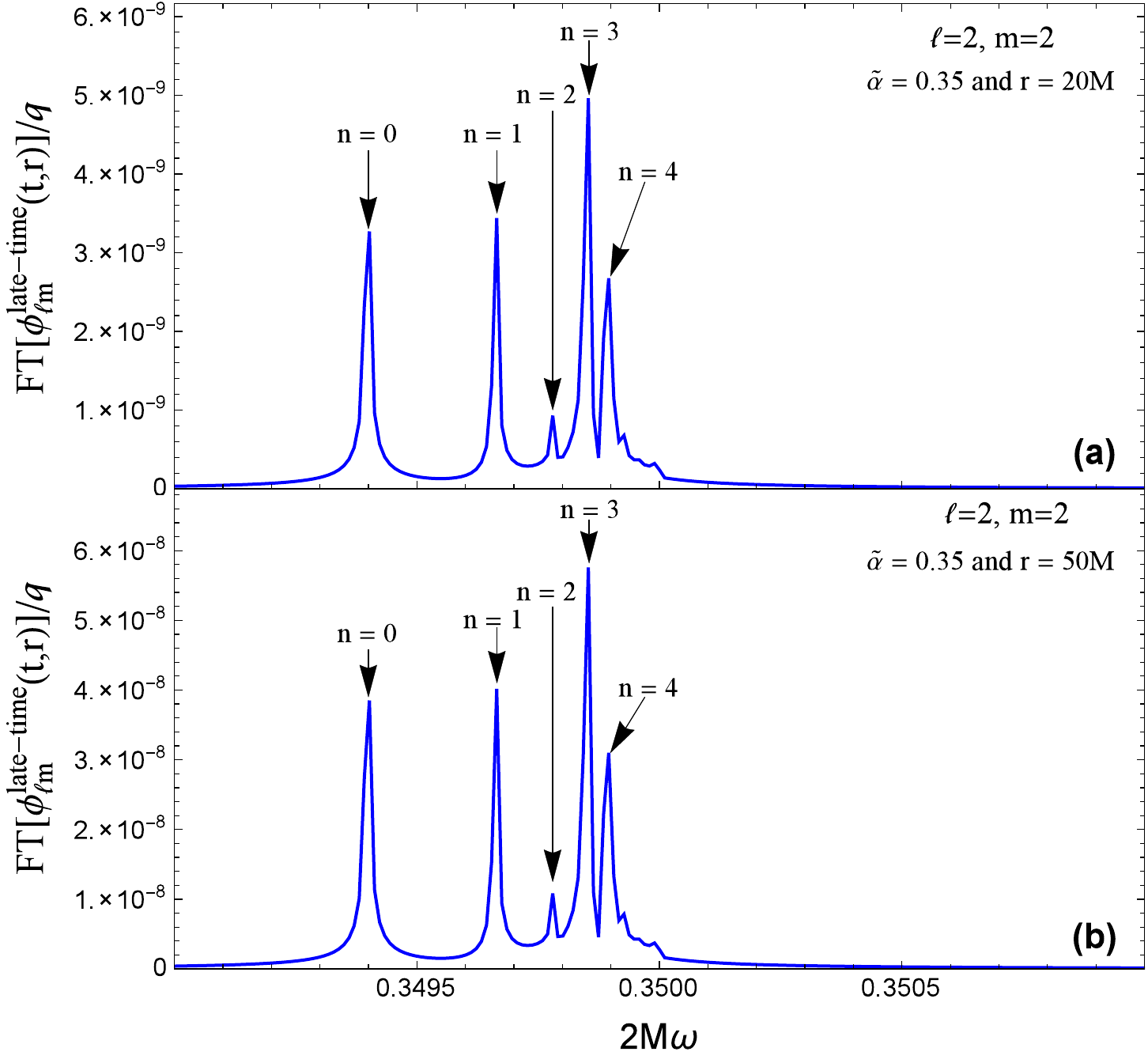}
           \vspace{-.3cm}
\caption{\label{QBS_2_0_mu_035}  Spectral content of the late-time phase of the quadrupolar waveform produced by the plunging particle. The results are obtained for a massive scalar field  ($\tilde\alpha=0.35$ is above the threshold $\tilde\alpha_c \approx0.2722 $) and the observer is located at (a) $r=20M$ and (b) $r=50M$. We observe the signature of the first long-lived QBSs.}
\end{figure}

In Figs.~\ref{QBS_2_0_mu_025} and \ref{QBS_2_0_mu_035}, we display the spectral content of the late-time phase of the quadrupolar waveform produced by the plunging particle. We can observe the excitation of QBSs (see also Table~\ref{tab:QBS}). We can also note that, as the reduced mass parameter $\tilde\alpha$ increases, the spectrum of the frequencies of the QBSs spreads more and more and it is then possible to separate the different excitation frequencies.

\section{Conclusion and perspectives}
\label{conclusion}

If the graviton has a mass, the study of the gravitational radiation generated by a particle plunging into a BH is certainly a problem of fundamental importance whose solution could help us to test the various massive gravity theories. Because this problem seems to us rather difficult, theoretically as well as numerically, we have chosen to work with a toy model where the massive spin-2 perturbations of the BH are replaced by a massive scalar field and we have considered that this field and the plunging particle are linearly coupled. In our opinion, the study of the scalar radiation generated by the plunging particle has permitted us to exhibit some features that should also be present in the solution of the analogous problem in massive gravity because they are directly associated with characteristics shared by the massive scalar and spin-2 fields such as (i) the propagating or evanescent nature of the partial modes and (ii) the existence of QBSs.

In this article, by limiting our study to the case of the Schwarzschild BH and to the $(\ell=2,m=2)$ mode of the scalar field, we have more particularly shown that:

\begin{itemize}[label={$-$}]

 \item The waveform produced by a particle plunging into a Schwarzschild BH from slightly below the ISCO can be roughly decomposed in three phases: (i) an adiabatic phase corresponding to the quasicircular motion of the particle near the ISCO, (ii) a ringdown phase due to the excitation of QNMs and (iii) a late-time phase.

 \item In the adiabatic phase, for an observer at a large distance from the BH, the behavior of the waveform depends on a threshold value $\tilde\alpha_{c}$ of the reduced mass parameter $\tilde\alpha$. This threshold corresponds to the mass parameter $\mu_c = 2 \Omega_{\mathrm{ISCO}}$ where $\Omega_{\mathrm{ISCO}}$ denotes the orbital frequency  of the particle moving on the ISCO. This threshold value separates the dispersive and the evanescent regimes. Above the threshold, for an observer at spatial infinity, the waveform amplitude vanishes. For a given distance $r$ of the observer, the waveform amplitude decreases as the mass increases and vanishes for large masses.

\item The ringdown phase (oscillations and damping) is very well described from the excitation of the first QNM, i.e. the least damped one.
\item In the late-time phase, whatever the mass parameter, we can observe the excitation of QBSs. In the adiabatic phase, the excitation of QBSs only occurs for masses above the threshold. These QBSs could dominate the signal at a large distance from the BH.

\item For large masses, the ringdown phase is modified by the excitation of QBSs which blur the QNM contribution.

\end{itemize}

It should be noted that, {\it mutatis mutandis}, the behaviors of an arbitrary ($\ell, m$) waveform and of the quadrupolar waveform generated by the plunging particle are quite similar. In general, the waveform amplitude in the adiabatic phase decreases as $\ell$ increases or $m$ decreases (see Figs.~\ref{Amp_Rep_Circ_mu_l_2_3_4_10M_50M_100M} and \ref{Amp_Rep_Circ_l_4_mu_10M}). However, it should be noted that this does not remain true around the threshold values for an observer at a large distance from the BH (see Fig.~\ref{Amp_Rep_Circ_mu_l_2_3_4_10M_50M_100M}).

We hope in the near future to extend this work to the massive spin-2 perturbations of the Schwarzschild BH and to consider higher values of the mass parameter in order to check if the intrinsic giant ringings predicted in our previous works \cite{Decanini:2014kha,Decanini:2014bwa} are generated in physical processes. With this aim in view, there remain a lot of theoretical and numerical difficulties to overcome.

\begin{acknowledgments}

We gratefully acknowledge Thibault Damour for drawing some years ago the attention of one of us (A. F.) to the plunge regime. We wish also to thank Andrei Belokogne for various discussions and the ``Collectivit\'e Territoriale de Corse'' for its support through the COMPA project.

\end{acknowledgments}

\appendix

\section{Waveforms produced by a scalar point particle living on the ISCO}
\label{appen_A}

In this appendix, we provide a simple closed-form expression for the emitted waveform when the particle lives on the ISCO and we analyze its behavior as the reduced mass parameter $\tilde \alpha$ increases and as the distance $r$ of the observer changes. These results are helpful in order to describe the adiabatic phase of the waveform emitted by a point particle on a plunge trajectory (see Sec.~\ref{sec_3_c}).

\subsection{Source due to a point particle on a circular orbit and associated waveform}
\label{appen_A_1}

Here, we assume that the particle moves on a circular orbit with radius $r_{0}$, i.e. we have $r_{p}(\tau)=r_{0}=\mathrm{Const.}$ in the geodesic equations (\ref{geodesic_system}). After integration, they provide for the angular momentum and the energy of the particle
\begin{equation}
\label{angular_momentum_circular}
\frac{L}{m_{0}}=\left(\frac{M r_{0}}{1-3M/r_{0}}\right)^{1/2}
\end{equation}
and
\begin{equation}
\label{energy_circular}
\frac{E}{m_{0}}=\frac{\left(1-2M/r_{0}\right)}{\left(1-3M/r_{0}\right)^{1/2}}.
\end{equation}
Moreover, we find that the angular coordinate $\varphi_{p}$ is given by
\begin{equation}
\label{geodesic_circular_1}
\varphi_{p}(\tau)=\left(\frac{M}{r_{0}^{3}\left(1-3M/r_{0}\right)}\right)^{1/2}\!\tau,
\end{equation}
if we use the proper time to describe the particle motion and by
\begin{equation}
\label{geodesic_circular_2}
\varphi_{p}(t)=\Omega\, t
\end{equation}
if we use the Schwarzschild time. In Eq.~(\ref{geodesic_circular_2}),
\begin{equation}
\label{Angular_velocity}
\Omega=\left(\frac{M}{r_{0}^{3}}\right)^{1/2}
\end{equation}
denotes the angular velocity of the particle (i.e. its orbital frequency).

After integration and by using, in particular, the change of variable $\tau \to \varphi_{p}(\tau)$, we can show that (\ref{Density_Source_1}) can be written in the form (\ref{source_decomposition}) with
 \begin{eqnarray}
\label{term_Source_circular}
\rho_{\omega \ell m}(r)=&&\frac{\sqrt{2\pi}q}{r_{0}}\left(1-\frac{3M}{r_{0}}\right)^{1/2}\nonumber\\
&&\times \delta(r-r_{0})\delta(\omega-m \Omega)Y_{\ell m}^{*}\left(\frac{\pi}{2},0\right).
\end{eqnarray}

The $(\ell,m)$ waveform produced by the particle orbiting the BH on a circular orbit with radius $r_0$ can now be obtained by substituting the source term (\ref{term_Source_circular}) in Eq.~(\ref{partial_response_def}). After integration, we have
\begin{eqnarray}
\label{partial_response_circular}
\phi_{\ell m}(t,r) =& & -\frac{q}{2\,i\,r_{0}}\left(1-\frac{3M}{r_{0}}\right)^{1/2}Y_{\ell m}^{*}\left(\frac{\pi}{2},0\right)\nonumber\\
& &\times\left.\frac{\phi_{\omega \ell}^{\mathrm{up}}(r)\,\phi_{\omega \ell}^{\mathrm{in}}(r_{0})}{\omega A_{\ell}^{(-)}(\omega)}
\,\,e^{-i \omega t}\right|_{\omega=m \Omega}{.}
\end{eqnarray}

\noindent It should be noted that (i) $\phi_{\ell -m}= (-1)^{m}\phi_{\ell m}^{\ast} $, so we can limit our study to the modes with $m\ge 0$, and (ii)  $\phi_{\ell m}=0$ for $\ell-m$ odd. This will simplify the discussions in the next subsection.

\subsection{Waveforms for a particle on the ISCO}
\label{appen_A_2}

We now specialize to the point particle living on the ISCO, i.e. we assume that $r_{0}=6M$ in the waveform (\ref{partial_response_circular}) and we note also that (\ref{Angular_velocity}) reduces to

\begin{equation}
\label{Angular_velocity_ISCO}
\Omega_{\mathrm{ISCO}}=\frac{1}{6\sqrt{6}M}.
\end{equation}

\noindent We intend to highlight both the influence of the mass $\mu$ of the scalar field and the distance $r$ of the observer and to show more particularly that the classical and  well-known results concerning the massless field cannot be naively extended to massive fields.

\begin{figure}[h]
\centering
\vspace{0.15cm}
\includegraphics[scale=0.5]{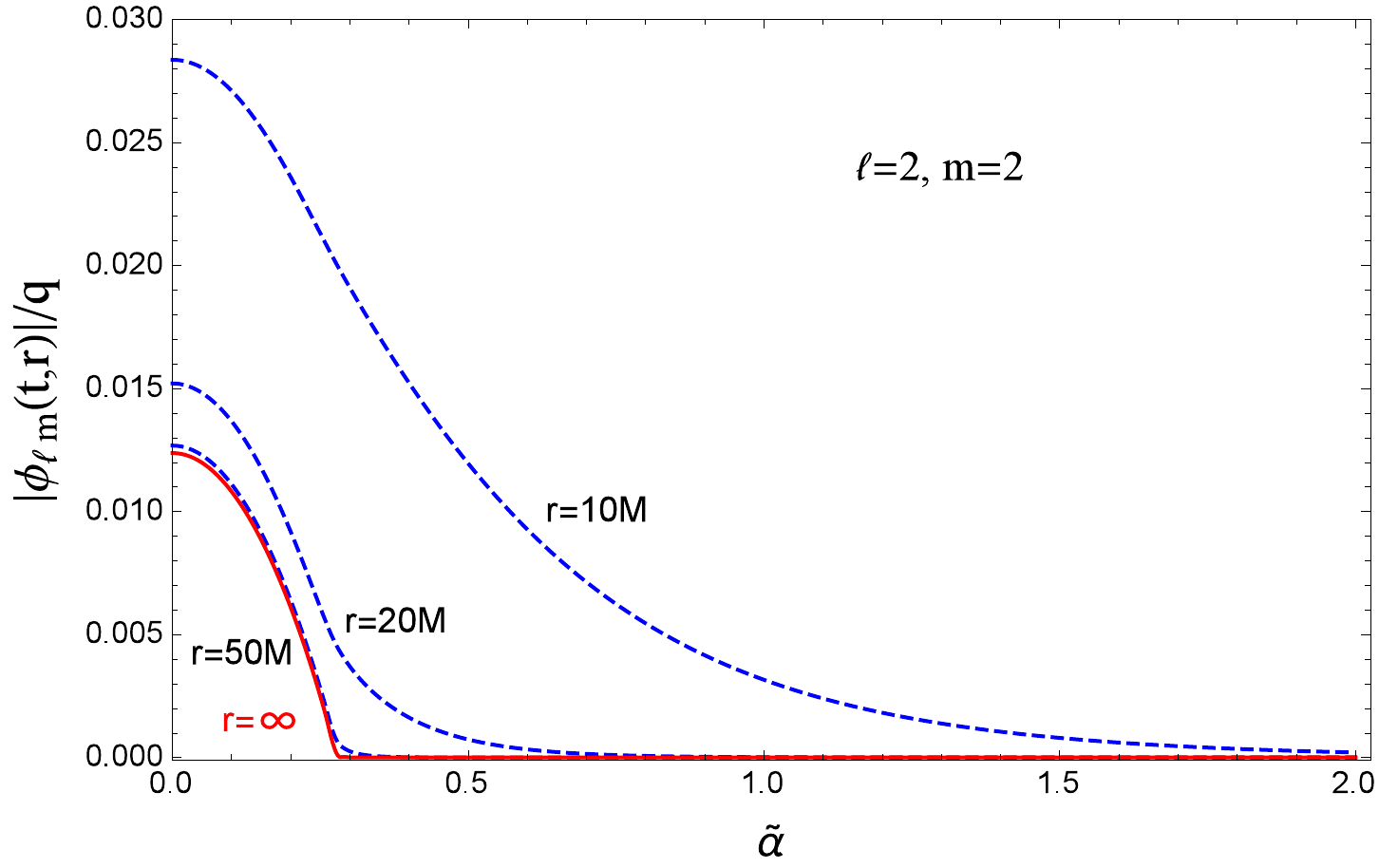}
\setlength\abovecaptionskip{0.55ex}
\vspace{-0.1cm}
\caption{\label{Amp_Rep_Circ_mu_l_2_10M_20M_200M} Waveform amplitudes for a particle on the ISCO and reduced masses in the range $\tilde\alpha \in [0, 2]$. The results are obtained from (\ref{partial_response_circular}) for the particular  $(\ell=2, m=2)$ mode. We study the role of the location of the observer. For an observer at large distance from the BH, the transition between the propagating (but dispersive) and evanescent behaviors of the response occurs at $\tilde\alpha_{c}\approx0.2722$.}
\end{figure}

In Fig.~\ref{Amp_Rep_Circ_mu_l_2_10M_20M_200M}, we focus on the $(\ell=2, m=2)$ waveform. For an observer at spatial infinity, the behavior of the waveform amplitude depends in a complex way on the mass of the scalar field. This is due to the term $p(\omega=m\Omega_{\mathrm{ISCO}})=\left[(m\Omega_{\mathrm{ISCO}})^{2}-\mu^{2}\right]^{1/2}$ which is implicitly present in the expression (\ref{partial_response_circular}) of the waveform. Indeed, depending on whether $\mu < \mu_{c}$  or $\mu > \mu_{c}$ with

\begin{equation}
\label{mu_c}
\mu_{c}=m\Omega_{\mathrm{ISCO}},
\end{equation}

\noindent this term is responsible for the propagating (but dispersive) or evanescent character of the $(\ell,m)$ mode of the massive field. In Fig.~\ref{Amp_Rep_Circ_mu_l_2_10M_20M_200M}, we observe the decreasing of the waveform amplitude when the reduced mass parameter $\tilde\alpha$ increases and we note that, above the threshold value $\tilde\alpha_c$ corresponding to $\mu_{c}$, the amplitude vanishes. As a consequence, above $\tilde\alpha_c$, the excitation of the system scalar field--BH by a particle orbiting on the ISCO cannot be observed at spatial infinity.  This abrupt behavior remains valid for large distances (for $r\ge 50M$) but is smoothed  for short distances with a vanishing of the amplitude for large masses. As a consequence, for ``very'' massive scalar fields, the excitation of the system scalar field--BH by a particle orbiting ISCO cannot be observed whatever the distance.

\begin{figure}[h]
\centering
       \includegraphics[scale=0.5]{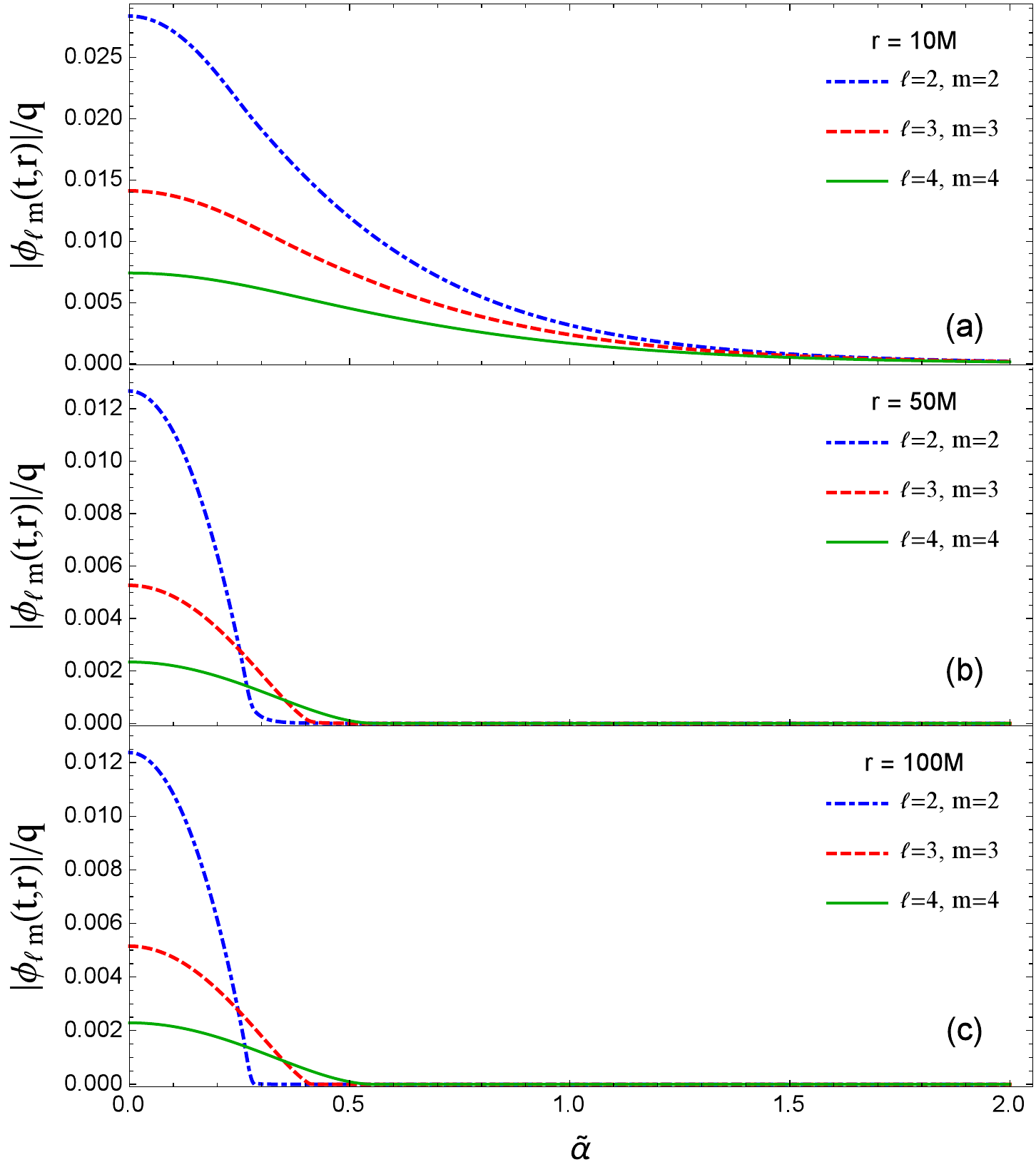}
          \vspace{-0.3cm}
\caption{\label{Amp_Rep_Circ_mu_l_2_3_4_10M_50M_100M} Waveform amplitudes for a particle on the ISCO and reduced masses in the range $\tilde\alpha \in [0, 2]$. The results are obtained from (\ref{partial_response_circular}) for some  $(\ell, m=\ell)$ modes and an observer at (a) $r=10M$, (b) $r=50M$ and (c) $r=100M$. The transition between the propagating (but dispersive) and evanescent behaviors of the response occurs at $\tilde\alpha_{c}\approx0.2722$ for the  $(\ell=2, m=2)$ mode, at $\tilde\alpha_{c}\approx0.4082$ for the $(\ell=3, m=3)$ mode  and at $\tilde\alpha_{c}\approx0.5443$ for the  $(\ell=4, m=4)$ mode.}
\end{figure}

\begin{figure}[h]
\centering
       \includegraphics[scale=0.5]{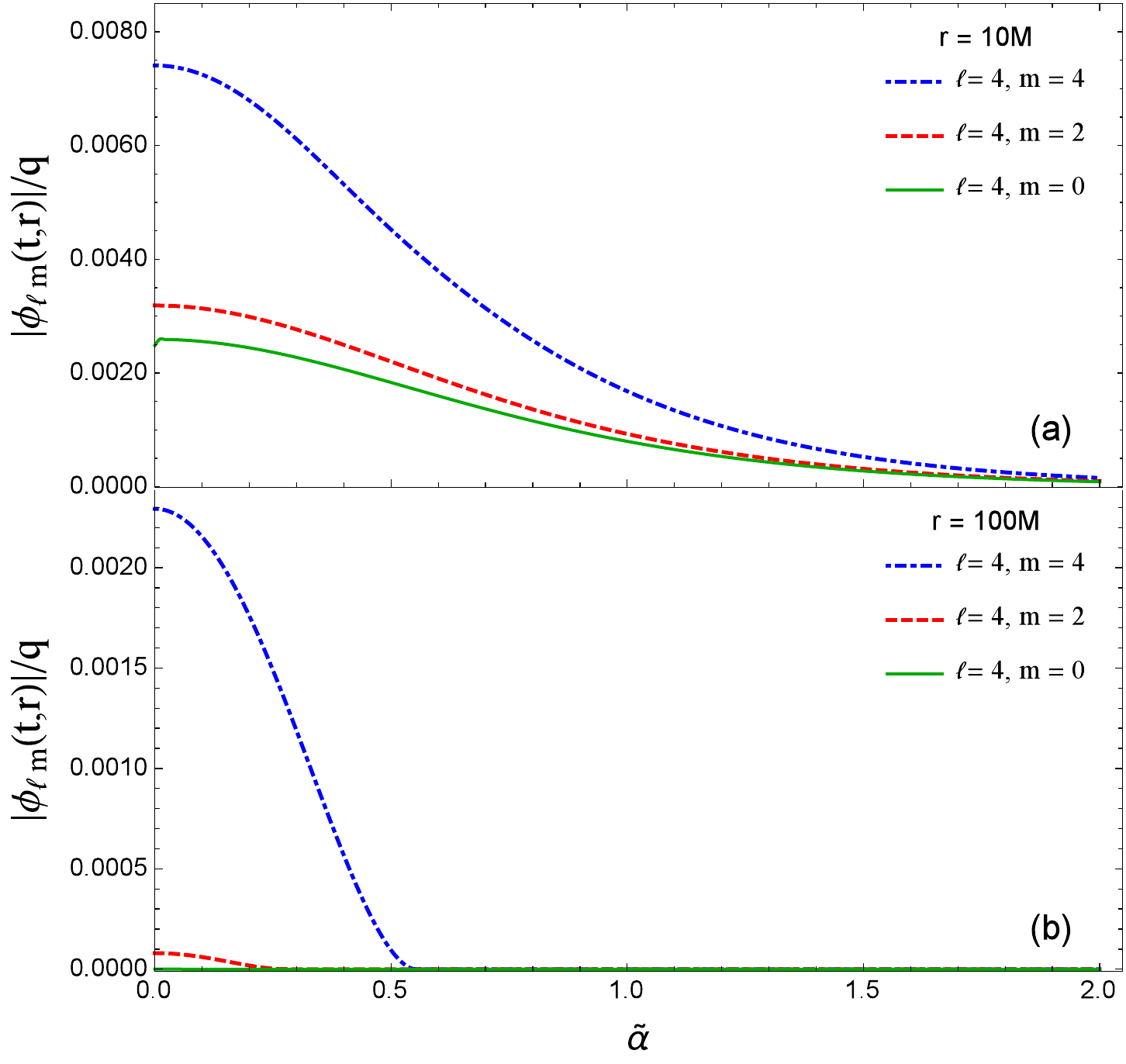}
          \vspace{-0.3cm}
\caption{\label{Amp_Rep_Circ_l_4_mu_10M} Waveform amplitudes for a particle on the ISCO and reduced masses in the range $\tilde\alpha \in [0, 2]$. The results are obtained  from (\ref{partial_response_circular}) for modes with angular momentum $\ell=4$. We study the influence of the azimuthal number $m$ for an observer at (a) $r=10M$ and (b) $r=100M$.}
\end{figure}

In Figs.~\ref{Amp_Rep_Circ_mu_l_2_3_4_10M_50M_100M} and  \ref{Amp_Rep_Circ_l_4_mu_10M}, we show that the behavior of the $(\ell=2, m=2)$ waveform can also be observed for arbitrary $(\ell, m)$ waveforms. Moreover, in Fig.~\ref{Amp_Rep_Circ_mu_l_2_3_4_10M_50M_100M}, it is interesting to note that, in a large domain around the particular threshold masses, the mode with the lowest $\ell$ does not provide the greatest waveform amplitudes. In this, a massive field is completely different from a massless one. Furthermore, in Fig.~\ref{Amp_Rep_Circ_l_4_mu_10M} we can observe that,  for a given $\ell$, the waveform amplitude decreases with $m$ and therefore that the $(\ell, m=\pm\ell)$ modes induce the highest amplitudes. Such behavior is well known for massless fields and we can see that it remains valid for massive fields.

\section{Quasinormal frequencies, QNMs and associated ringings}
\label{appen_B}

In this appendix, we consider the weakest damped QNM with $\ell=2$ and we obtain numerically its response when it is excited by the plunging particle. This result is helpful in order to describe the ringdown phase of the waveform emitted by a point particle on a plunge trajectory (see Sec.~\ref{sec_3_d}).

The response of a QNM to an excitation (the so-called quasinormal waveform or, in other words the associated ringing) can be constructed from (i) the intrinsic characteristics of the BH (the quasinormal frequency $\omega_{\ell n}$, the QNM itself and the corresponding excitation factor ${\cal{B}}_{\ell n}$ -- here $n=0$ corresponds to the fundamental QNM, i.e. the least damped one and $n=1, 2, \ldots$ to the overtones) and (ii) the source of the excitation. In the following subsections, we describe this construction.

\subsection{Quasinormal frequencies and excitation factors}
\label{appen_B_1}

\begin{figure}[h!]
\centering
       \includegraphics[scale=0.5]{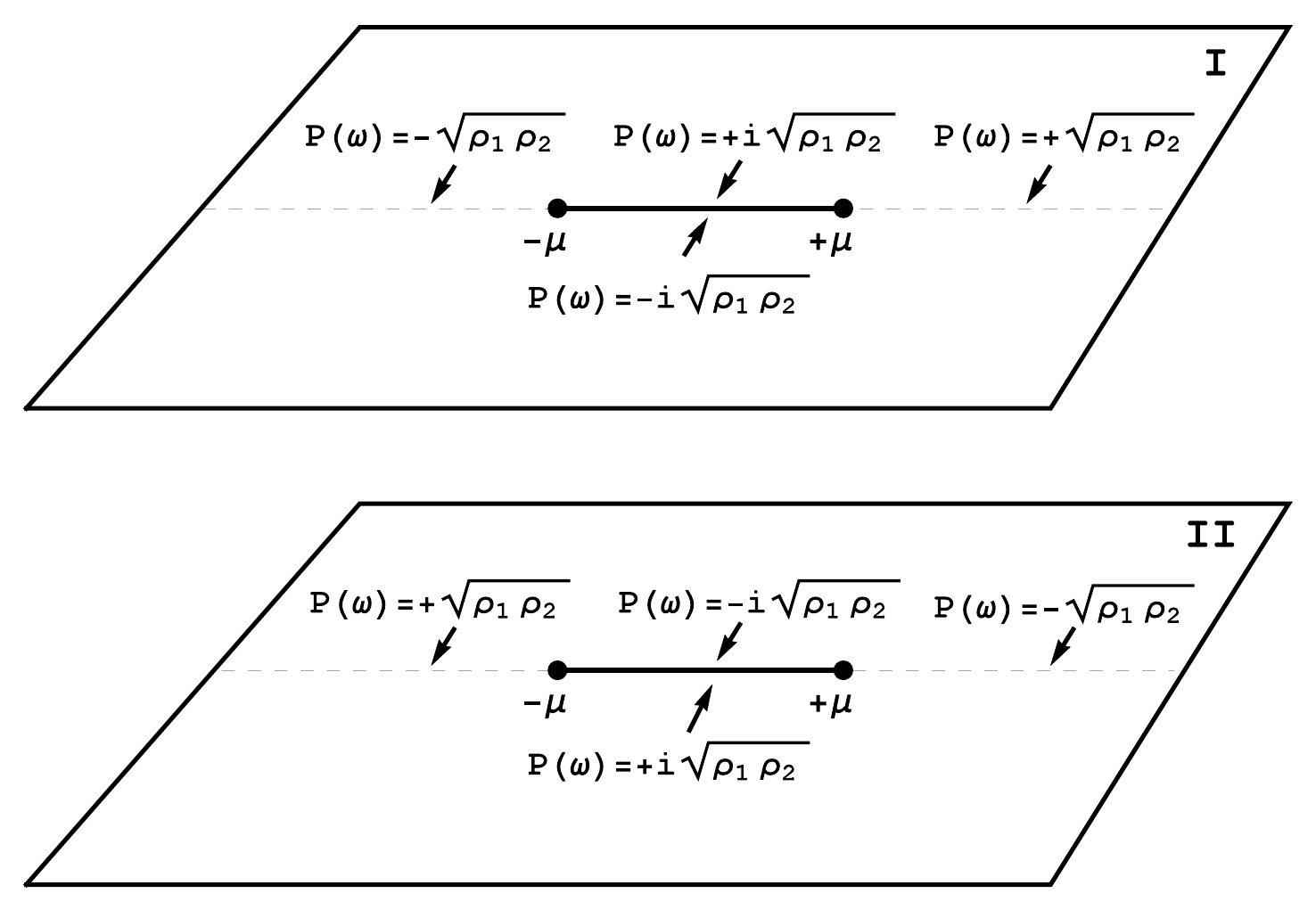}
\caption{\label{feuillets_QNM_QBS} The two Riemann sheets of the function $p(\omega)=(\omega^2-\mu^2)^{1/2}$ and the branch cut we have chosen. We have introduced the polar coordinates  $(\rho_1,\theta_1)$ and $(\rho_2,\theta_2)$ such that $\omega-\mu =\rho_1 e^{i\theta_1}$ and $\omega+\mu =\rho_2 e^{i\theta_2}$. Here $0\leq\theta_1,\theta_2<2\pi$.}
\end{figure}

We recall that the quasinormal frequencies $\omega_{\ell n}$ are the zeros of the Wronskian $W_{\ell}(\omega)$ given by Eq.~(\ref{Well}) lying in the lower part of the first Riemann sheet associated with the function $p(\omega)=(\omega^2-\mu^2)^{1/2}$ (see Fig.~\ref{feuillets_QNM_QBS}) and that the corresponding excitation factors are defined by

\begin{equation}
\label{excitation_factor}
{\cal{B}}_{\ell n}=\left[\frac{1}{2p(\omega)}\,\,\frac{A_{\ell}^{(+)}(\omega)}{\frac{d A_{\ell}^{(-)}(\omega)}{d \omega}}\right]_{\omega=\omega_{\ell n}}.
\end{equation}

\noindent It should be noted that when the Wronskian $W_{\ell}(\omega)$ vanishes (i.e. if $\omega=\omega_{\ell n}$), the functions $\phi_{\omega\ell}^{\mathrm{in}}$ and  $\phi_{\omega\ell}^{\mathrm{up}}$ are linearly dependent and propagate inward at the horizon and outward at spatial infinity, such behavior defining the QNMs.

We also recall that the complex spectrum of the quasinormal frequencies is symmetric with respect to the imaginary $\omega$ axis. In other words, if $\omega_{\ell n}$ is a quasinormal frequency lying in the fourth quadrant, $-\omega_{\ell n}^{*}$ is the symmetric quasinormal frequency lying in the third one. In Table~\ref{tab:QNM}, we have considered the least damped $\ell =2$ QNM and we have given its quasinormal frequency $\omega_{2 0}$ with positive real part and the associated  excitation factor ${\cal{B}}_{2 0}$ for three particular values of the reduced mass parameter $\tilde\alpha$. The results have been obtained by using the numerical methods described in Sec.~II of Ref.~\cite{Decanini:2014bwa}.

\begin{table}[h]
\caption{\label{tab:QNM} Evolution of the quasinormal frequency and the excitation factor of the fundamental $\ell=2$ QNM.}
\smallskip
\centering
\begin{tabular}{lcccccr}
\hline
\hline
$(\ell,n)$  &${\tilde \alpha}$ & $2M \omega_{\ell n}$& ${\cal{B}}_{\ell n}$\\
\hline
$(2,0)$ & $0$  & $0.96729\, -0.19352 i$  & $0.11935\, + 0.01343 i$ \\
\hline
$(2,0)$ & $0.25$  & $0.97717\, -0.19012 i$  & $0.12241\, + 0.00737 i$ \\
\hline
$(2,0)$ & $0.35$  & $0.98669\, -0.18684 i$  & $0.12548\, + 0.00091 i$\\
\hline
\hline
\end{tabular}%
\end{table}

\subsection{Quasinormal waveforms}
\label{appen_B_2}

We now deform the contour of integration in Eq.~(\ref{partial_response_def}) in order to extract a residue series over the quasinormal frequencies. It is given by

\begin{equation}
\label{phi_QNM_1}
\phi_{\ell m}^{\mathrm{QNM}}(t,r)=\sum^{+\infty}_{n=0}\phi_{\ell m n}^{\mathrm{QNM}}(t,r)
\end{equation}

\noindent with

\begin{eqnarray}
\label{phi_QNM_2}
\phi_{\ell m n}^{\mathrm{QNM}}(t,r)=&&\sqrt{2\pi}\left({\cal{C}}_{\ell m n}\,\,\phi_{\omega_{\ell n}\ell}^{\mathrm{up}}(r_{\ast})e^{-i \omega_{\ell n}t}\,\,\vphantom{e^{i \omega_{\ell n}^{\ast}t}}\right.\nonumber\\
&&\left.+\,\,{\cal{D}}_{\ell m n}\left[\phi_{\omega_{\ell n}\ell}^{\mathrm{up}}(r_{\ast})\right]^{\ast}e^{i \omega_{\ell n}^{\ast}t}\right).
\end{eqnarray}

\noindent Here ${\cal{C}}_{\ell m n}$ and ${\cal{D}}_{\ell m n}$ denote the extrinsic excitation coefficients defined by

\begin{equation}
\label{excitation_coeff_C}
{\cal{C}}_{\ell m n}={\cal{B}}_{\ell n}\left[\int_{2M}^{+\infty} dr' \, \frac{\phi_{\omega \ell}^{\mathrm{in}}(r')\rho_{\omega\ell m}(r')}{\left[\omega/p(\omega)\right]\, A_{\ell}^{(+)}(\omega)} \right]_{\omega=\omega_{\ell n}}
\end{equation}

\noindent and

\begin{equation}
\label{excitation_coeff_D}
{\cal{D}}_{\ell m n}=[{\cal{B}}_{\ell n}]^{*}\left[\int_{2M}^{+\infty}\!\!\! dr' \, \frac{\phi_{\omega \ell}^{\mathrm{in}}(r')\rho_{\omega\ell m}(r')}{\left[\omega/p(\omega)\right]\, A_{\ell}^{(+)}(\omega)} \right]_{\omega=-\omega_{\ell n}^{*}}\!\!\!\!\!\!\!\!\!\!\!\!\!\!\!\!\!\!.
\end{equation}

\noindent The first term in Eq.~(\ref{phi_QNM_2}) is the contribution of the quasinormal frequency $\omega_{\ell n}$ lying in the fourth quadrant of the first Riemann sheet of $p(\omega)$ while the second one is the contribution of $-\omega_{\ell n}^{\ast}$, i.e. its symmetric with respect to the imaginary axis. The expression (\ref{phi_QNM_2}) has been simplified by using some symmetry properties in the change $\omega_{\ell n}\rightarrow-\omega_{\ell n}^{\ast}$ and, in particular,

\begin{subequations}
\begin{eqnarray}
&&p(-\omega_{\ell n}^{\ast})=-[p(\omega_{\ell n})]^{\ast},\\
&&\phi_{-\omega_{\ell n}^{\ast} \ell}^{\mathrm{in}}=[\phi_{\omega_{\ell n} \ell}^{\mathrm{in}}]^{\ast},\\
&&\phi_{-\omega_{\ell n}^{\ast} \ell}^{\mathrm{up}}=[\phi_{\omega_{\ell n} \ell}^{\mathrm{up}}]^{\ast}.
\end{eqnarray}
\end{subequations}

\noindent It should be noted that, in our problem, the spherical symmetry of the Schwarzschild BH is broken due to the asymmetric plunging trajectory. It is this dissymmetry which, in connection with the presence of the azimuthal number $m$, forbids us to gather the two terms in Eq.~(\ref{phi_QNM_2}).

Let us finally remark that $\phi_{\ell m n}^{\mathrm{QNM}}(t,r)$ does not provide physically relevant results at ``early times'' due to its  exponentially divergent behavior as $t$ decreases. It is necessary to determine, from physical considerations when this is possible or by using a numerical approach, the time beyond which this waveform can be used, i.e. the starting time $t_\mathrm{start}$ of the BH ringing.

\section{Complex frequencies of the first QBSs}
\label{appen_C}

The complex frequencies $\omega_{\ell n}$ of the QBSs are the zeros of the Wronskian $W_{\ell}(\omega)$ given by Eq.~(\ref{Well}) lying in the lower part of the second Riemann sheet associated with the function $p(\omega)=(\omega^2-\mu^2)^{1/2}$ (see Fig.~\ref{feuillets_QNM_QBS}). Their  spectrum is symmetric with respect to the imaginary $\omega$ axis. They can be numerically obtained by using the method which has permitted us to determine in Sec.~\ref{appen_B_1} the quasinormal frequencies but now, because we are working on the second Riemann sheet associated with $p(\omega)$, it is necessary to use $-p(\omega)$ instead of $p(\omega)$ (see also Fig.~\ref{feuillets_QNM_QBS}). In Table~\ref{tab:QBS}, we  have given a sample of the complex frequencies of the long-lived QBSs relevant to the spectral content of the waveform emitted by the plunging particle (see Sec.~\ref{sec_3_e}).

\begin{table}[h]
\caption{\label{tab:QBS} Complex frequencies $\omega_{\ell n}$ of the first long-lived QBSs.}
\smallskip
\centering
\begin{tabular}{lcccccr}
\hline
\hline
& $(\ell,n)$  &$ {\tilde \alpha}$ & $2M \omega_{\ell n}$  &\\
\hline
& $(2,0)$ & $0$  &   -----    &\\
\hline
& $(2,0)$ & $0.25$  & $0.24978\, -6.08009\times10^{-17} i$  &\\
& $(2,1)$ &         & $0.24988\, -3.63454\times10^{-17} i$  &\\
\hline
& $(2,0)$ & $0.35$  & $0.34940\, -1.08832\times10^{-14} i$  &\\
& $(2,1)$ &         & $0.34966\, -1.03797\times10^{-14} i$  &\\
& $(2,2)$ &         & $0.34978\, -3.90581\times10^{-15} i$  &\\
& $(2,3)$ &         & $0.34985\, -2.42608\times10^{-15} i$  &\\
& $(2,4)$ &         & $0.34989\, -1.59115\times10^{-15} i$  &\\
\hline
\hline
\end{tabular}%
\end{table}

\bibliography{Waveforms_Massive_Scalar_Field}

\end{document}